\documentclass[10pt, journal, final]{IEEEtran}

\usepackage[letterpaper, top=1.9cm, bottom=2.54cm, left=1.6cm, right=1.6cm]{geometry}
\usepackage[utf8]{inputenc}
\usepackage[T1]{fontenc}
\usepackage{ifthen}
\usepackage[cmex10]{amsmath} 

\usepackage{amsmath}
\usepackage{amssymb}
\usepackage{amsfonts}
\usepackage{algorithm}
\usepackage{algpseudocode}
\usepackage{comment}
\usepackage{dsfont}
\usepackage{float}
\usepackage{cite}
\usepackage{graphicx}
\usepackage{epsfig}
\usepackage{subfigure}
\usepackage{psfrag}
\usepackage{xcolor}
\usepackage{url}

\usepackage{bm}
\allowdisplaybreaks
\usepackage[colorlinks,linkcolor=black,urlcolor=black,anchorcolor=black,citecolor=black,hyperfootnotes=true]{hyperref}
\newtheorem{theorem}{{Theorem}}
\newtheorem{remark}{Remark}
\newtheorem{proposition}{Proposition}

\begin{document}
\abovedisplayskip=1pt
\belowdisplayskip=1pt
\allowdisplaybreaks
	
\title{Joint Transmission and Compression Optimization for Networked Sensing with Limited-Capacity Fronthaul Links}

\author{
Weifeng~Zhu, 
Shuowen~Zhang,
and Liang~Liu,

\thanks{
This paper was presented in part at the IEEE Global Conference of Communications (GLOBECOM) 2024 \cite{Zhu_2024_GC}.}
\thanks{The authors are with the Department of Electrical and Electronic Engineering, The Hong Kong Polytechnic University, Hong Kong SAR (e-mail: \{eee-wf.zhu, shuowen.zhang, liang-eie.liu\}@polyu.edu.hk).}
}


\maketitle

\begin{abstract}
This paper considers networked sensing in cellular network, where multiple base stations (BSs) first compress their received echo signals from multiple targets and then forward the quantized signals to the central unit (CU) via limited-capacity fronthaul links, such that the CU can leverage all useful echo signals to perform high-resolution localization. Under this setup, we manage to characterize the posterior Cramér-Rao Bound (PCRB) for localizing all the targets with random positions as a function of the transmit covariance matrix and the compression noise covariance matrix of each BS. Then, a PCRB minimization problem subject to the transmit power constraints and the fronthaul capacity constraints is formulated to jointly design the BSs' transmission and compression strategies. We propose an efficient algorithm to solve this problem based on the alternating optimization technique. Specifically, it is shown that when either the transmit covariance matrices or the compression noise covariance matrices are fixed, the successive convex approximation (SCA) technique can be leveraged to optimize the other type of covariance matrices locally optimally. Moreover, we also propose a novel estimate-then-beamform-then-compress strategy for the massive receive antenna scenario, under which each BS first estimates targets' angle-of-arrivals (AOAs) locally, then beamforms its high-dimension received signals into low-dimension ones based on the estimated AOAs, and last compresses the beamformed signals for fronthaul transmission. An efficient beamforming and compression design method is devised under this strategy. Numerical results are provided to verify the effectiveness of our proposed algorithms.
\end{abstract}

\section{Introduction}

Integrated sensing and communication (ISAC) is identified as one of the six key usage scenarios of the six-generation (6G) cellular systems \cite{ITU_2023}.
The main challenge in ISAC is how to embed the sensing functionality into the future 6G communication systems efficiently \cite{isac_survey1,isac_survey2}. In the literature, plenty of works have studied the practical strategies to perform sensing \cite{Rahman_2020_TAES,Zhang_2021_JSTSP,Sohrabi_2022_JSAC} and the fundamental limits of sensing \cite{Shen_2010_TIT,Xiong_2023_TIT,Liu_2022_CST} in communication systems. It is worth noting that in most of the existing ISAC works, the sensing functionality is achieved via one base station (BS), just as in radar systems.

However, there exists a notable difference between the BSs and the conventional radars - the BSs across large-scale area are inter-connected via the fronthaul links and can share information with each other. Actually, this property has been utilized to achieve cooperative communication in cellular network, giving rise to cloud radio access network (C-RAN), cell-free massive multiple-input multiple-output (MIMO), etc., where multiple BSs jointly serve the users via sharing the data messages and the channel state information (CSI). Similarly, in the future 6G ISAC systems, multiple BSs may share the local sensing information with each other to perform networked sensing \cite{Liu_2024_arxiv}. This is a major advantage of 6G sensing systems over the conventional radar systems.

Motivated by the cooperation gain in communication, this paper considers the networked sensing technique, under which multiple BSs forward their received echo signals reflected by multiple targets to a central unit (CU), and the CU jointly localizes all the targets based on the global echo signals. Theoretically speaking, the sensing performance can be significantly enhanced by fusing the global sensing information at the CU. However, in practice, the BSs are connected to the CU via limited-capacity fronthaul links and they can only transmit compression bits over these links. Therefore, the echo signals received by the CU are subject to compression noise, whose power depends on the compression strategy adopted by each BS. This indicates that besides the transmission strategy that is considered in single-BS based sensing scenario, each BS should also devise its compression strategy carefully under networked sensing. 
In this paper, we make an early attempt to jointly design the transmission and compression strategies of each BS under networked sensing with limited-capacity fronthaul.

\subsection{Related Works}

The functionality of sensing is already recognized to be indispensable in the future 6G wireless systems and has attracted many interests of researchers.
Under the single-BS sensing setup, previous works have investigated the fundamental limits of sensing \cite{Bek_2006_TSP,Li_2008_TSP,Wilcox_2012_TSP,Liu_2022_TSP,Wang_2023_TWC,Hua_2024_TWC,Xu_2024_JSAC,Yao_2024_arxiv,Hou_2024_arxiv,Attiah_2024_ISIT} to provide useful insights for the practical sensing technique design. Therein, the works \cite{Bek_2006_TSP,Li_2008_TSP,Wilcox_2012_TSP,Liu_2022_TSP,Wang_2023_TWC,Hua_2024_TWC} derive the Cramér-Rao Bound (CRB) to characterize the localization performance with unknown but deterministic target locations. Then, the works \cite{Xu_2024_JSAC,Hou_2024_arxiv,Yao_2024_arxiv,Attiah_2024_ISIT} consider the scenario where the target locations are random with known distribution and give the explicit expression of the posterior Cramér-Rao Bound (PCRB) on target directions by exploiting the location distribution. By regarding the CRB or the PCRB as the objective function, the works \cite{Li_2008_TSP,Wilcox_2012_TSP,Liu_2022_TSP,Wang_2023_TWC,Hua_2024_TWC,Xu_2024_JSAC,Yao_2024_arxiv,Attiah_2024_ISIT,Hou_2024_arxiv} propose to optimize the waveform to realize the optimal sensing performance under certain constraints. 
Besides the single-BS setup, the performance limits in terms of the CRB for target localization under networked sensing are also studied in \cite{Bar_2022_TAES,Gao_2023_JSAC}. Specifically, the work \cite{Bar_2022_TAES} considers the system where both the transmitters and receivers are equipped with one single antenna and analyzes the CRB for wideband localization.
Then, the work \cite{Gao_2023_JSAC} derives the CRB for the cooperative ISAC system where multiple BSs localizes the targets by estimating their directions and goes forward to optimize the transmit beamformers to balance the trade-off between sensing CRB and transmission rate. However, both of these two works do not take the practical limited-capacity fronthaul issue in cooperative networks \cite{Simenone_2016_JCN,Liu_2015_TSP,Zhou_2016_TIT} into consideration. In networked sensing systems, the transmit strategy and the compression strategy of the BSs should be jointly optimized.

Several works make efforts in studying the fundamentals of networked sensing by considering the effect of compression arising from fronthaul links. 
The work \cite{Khalili_2015_SPL} adopts the information-theoretic criterion of the Bhattacharyya distance and quantization rate to measure the detection performance and backhaul capacity, respectively, in the cloud radio-multistatic radar system, which are then used for the joint optimization of the temporal code vector and backhaul quantization. 
Furthermore, the work \cite{Li_2023_VTC} derives the expected CRB for cooperative localization and considers the optimization of the bit allocation among the receive BSs.
However, the works \cite{Khalili_2015_SPL,Li_2023_VTC} consider the case of one transmit BS and multiple receive BSs, with each BS equipped with one antenna. In practical multi-cell communication systems, multiple BSs will transmit at the same time, resulting in inter-cell echo signals for localization. Such signals can make the characterization of localization PCRB quite challenging, which is not considered in the above works. Moreover, current BSs are equipped with multiple antennas. On one hand, targets' angle information can be extracted with multi-antenna receive BSs, which is not considered in \cite{Khalili_2015_SPL,Li_2023_VTC}. On the other hand, due to the multiple antennas at the BSs, the transmit beamforming strategy should be jointly optimized with the receive compression strategy, which is not considered in \cite{Khalili_2015_SPL,Li_2023_VTC}. 

\begin{figure}
\centering
\includegraphics[scale=0.4]{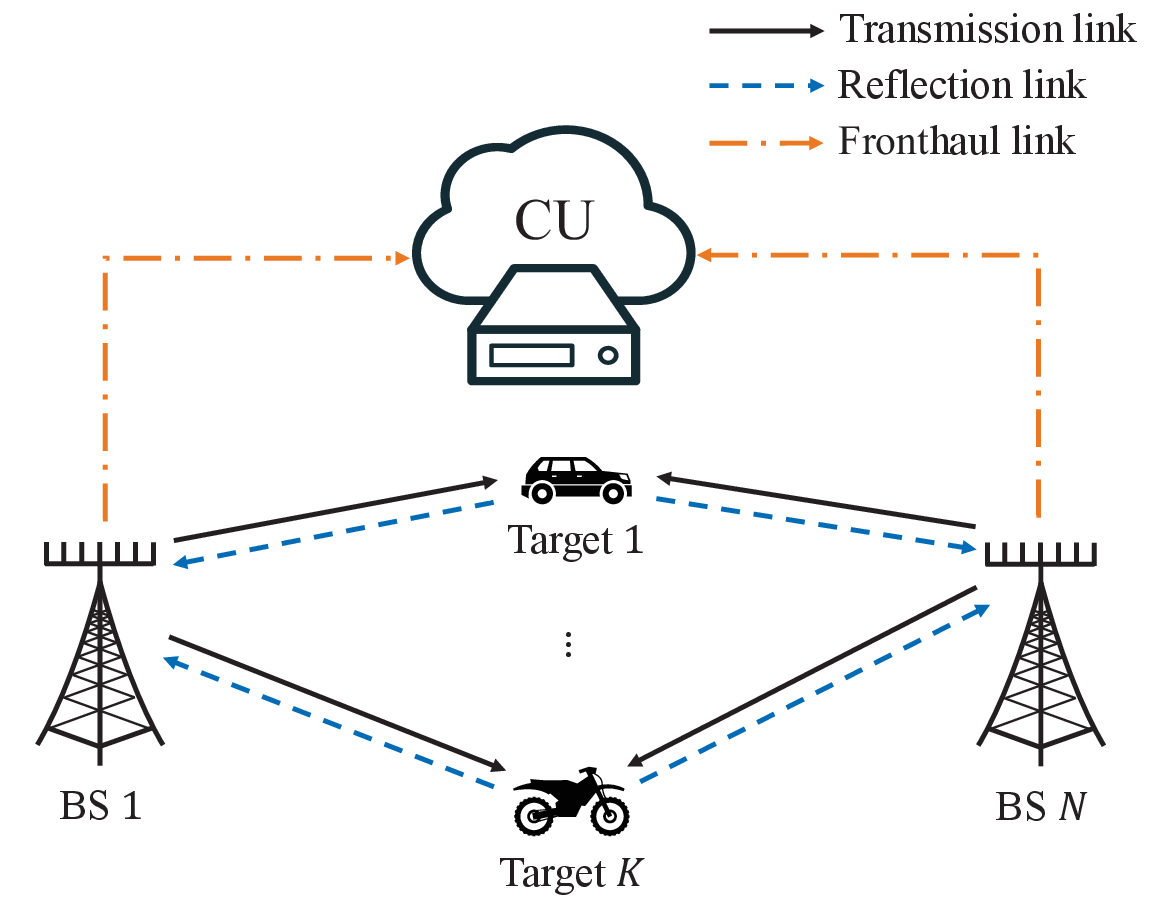}
\caption{System model for networked sensing with limited-capacity fronthaul.}\vspace{-10pt}
\label{Fig:SM}
\end{figure}

\subsection{Main Contributions}

In this paper, we consider a 6G networked sensing system consisting of multiple BSs, each equipped with multiple antennas, and multiple targets, as shown in Fig. \ref{Fig:SM}. Each BS first emits wireless signals in the downlink and then receives the echo signals from the targets. Next, each BS adopts a compress-then-forward strategy, under which it first compresses its received echo signals and then forwards the compression bits to the CU via a limited-capacity fronthaul link. At last, the CU may utilize the global information observed by all the BSs to jointly localize the targets. 
Our main contributions are summarized as follows:
\begin{itemize}
    \item Under the networked sensing setup with multiple BSs equipped with multiple antennas and multiple randomly located targets, we manage to characterize the PRCB for localization as a function of the BSs' transmit covariance matrices and compression noise covariance matrices. Such a result can be viewed as a generalization of the classic result under \cite{Li_2008_TSP} which characterizes the CRB with a single anchor. 
    \item Next, we formulate a joint transmission and compression optimization problem, which minimizes the PCRB subject to each BS's transmit power constraint and fronthaul capacity constraint. The formulated problem is non-convex and quite challenging to solve, due to the coupling of the transmit covariance matrices and compression noise covariance matrices. To deal with the these issues, we first propose an alternating optimization (AO)-based method to separately optimize the transmit covariance matrices and the compression noise covariance matrices. Moreover, we show that when each type of covariance matrices is fixed, we can apply the successive convex approximation (SCA) technique to optimize the other type locally optimally.
    \item We also propose an estimate-then-beamform-then-compress (EBC) strategy to reduce the algorithm complexity when the number of receive antennas is large. Specifically, under the proposed EBC strategy, each BS first estimates the angle-of-arrivals (AOAs) of the targets, and then exploits the AOA information to beamform the high-dimension received signals onto reduced dimensions, each corresponding to the main AOA of some target, and last compresses the lower-dimension beamformed signals for fronthaul transmission. This method is shown numerically to achieve very attractive performance with significantly reduced computational complexity.
\end{itemize}

\subsection{Organizations and Notations}
The rest of the paper is organized as follows. Section II gives the introduction of the system model and Section III derives the expression of the PCRB for localization in the considered system. In Section IV, the problem of joint transmission and compression optimization is formulated and the AO-based algorithm is proposed to solve it. Then an EBC strategy is introduced in Section V. Finally, Section VI evaluates the performance of the proposed algorithms and Section VII concludes this paper.

In this paper, vectors and matrices are denoted by boldface lower-case letters and boldface upper-case letters, respectively. For a complex vector $\boldsymbol{x}$, $||\boldsymbol{x}||_q$ and $x_n$ denote the $l_q$-norm and the $n$-th element, respectively. For an $M \times N$ matrix $\boldsymbol{X}$, $\boldsymbol{X}^T$, $\boldsymbol{X}^{*}$, $\boldsymbol{X}^{H}$, and $\boldsymbol{X}(m,n)$ denote its transpose, conjugate, conjugate transpose and element located at the $m$th row and $n$th column, respectively. For a square matrix $\boldsymbol{A}$, $|\boldsymbol{A}|$, $\text{tr}(\boldsymbol{A})$, and $\boldsymbol{A}^{-1}$ denote its determinant, trace, and inverse, respectively, and $\boldsymbol{A} \succeq \boldsymbol{0}$ means that $\boldsymbol{A}$ is positive semidefinite. The distribution of a circularly symmetric complex Gaussian random vector with mean $\boldsymbol{\mu}$ and covariance $\boldsymbol{\Sigma}$ is denoted by $\mathcal{CN}(\boldsymbol{\mu},\boldsymbol{\Sigma})$. Further, $\mathbb{E}_{\boldsymbol{a}}[\cdot]$ denotes the expectation operation over random vector $\boldsymbol{a}$, operation $\odot$ denotes Hadamard product of two matrices, and $\mathcal{R}\{\cdot\}$ denote the column space of the argument. Letters $\mathbb{R}^{M \times N}$ and $\mathbb{C}^{M \times N}$ denote the region of real matrix and complex matrix, respectively, which have the size of $M \times N$. Finally, letters $\mathbb{S}^{N}_{+}$ and $\mathbb{S}^{N}_{++}$ denote the region of positive semidefinite matrices and positive definite matrices, respectively, with size $N \times N$.

\section{System Model}

We consider the sensing job in a multi-cell ISAC system as shown in Fig. \ref{Fig:SM}, where $N$ BSs, each equipped with $M_t$ transmit antennas and $M_r$ receive antennas, are connected to the CU via the fronthaul links and cooperatively localize $K$ targets via utilizing the wireless signals. We focus on a two-dimensional (2D) coordinate system, while the known position of BS $n$ is denoted as $\boldsymbol{\rho}_n = [\rho_n^x,\rho_n^y]^T \in \mathbb{R}^{2 \times 1}$, and the unknown position of target $k$ is denoted as $\boldsymbol{q}_k = [q_k^x,q_k^y]^T \in \mathbb{R}^{2 \times 1}$, $n=1,\ldots,N$ and $k=1,\ldots,K$. In this paper, we assume that the unknown target positions $\boldsymbol{q} = [\boldsymbol{q}_1^T,\dots,\boldsymbol{q}_{K}^{T}]^T \in \mathbb{R}^{2K \times 1}$ are random with a joint probability density function (PDF) of $p_{\boldsymbol{q}}(\boldsymbol{q})$. Moreover, the angle-of-departure (AOD) of the signal transmitted from BS $n$ to target $k$ and the angle-of-arrival (AOA) of the signal reflected from target $k$ to BS $n$ are denoted as
\begin{equation}\label{equ:theta_xy}
    \theta_{n,k} = \arctan\frac{q_k^x - \rho_n^x}{q_k^y - \rho_{n}^y}, \quad\forall n,k.
\end{equation}
For convenience, in the rest of this paper, we define $\theta_{n,k}$ as the angle between BS $n$ and target $k$. Moreover, define $\boldsymbol{v}(\theta_{n,k}) \in \mathbb{C}^{M_t \times 1}$ and $\boldsymbol{a}(\theta_{n,k}) \in \mathbb{C}^{M_r \times 1}$ as the transmit and receive steering vectors of BS $n$ towards target $k$, $\forall n,k$.

Under the above system, all the BSs transmit wireless signals simultaneously in the downlink, then receive the echo signals from the targets, and forward their received echo signals to the CU via their fronthaul links for performing localization. Specifically, let $\boldsymbol{x}_n \in \mathbb{C}^{M_t \times 1}$ with the covariance matrix $\boldsymbol{R}_n = \mathbb{E}[\boldsymbol{x}_n\boldsymbol{x}_n^H]$ denote the signal emitted by the transmit antenna array of BS $n$, $\forall n$. Then, the echo signals at the receive antenna array of BS $n$ can be written as
\begin{align}\label{equ:signal_model}
    \boldsymbol{y}_{n} & = \boldsymbol{A}_{n}\sum_{u=1}^{N}  \boldsymbol{B}_{n,u} \boldsymbol{V}_{u}^T \boldsymbol{x}_{u} + \boldsymbol{z}_{n} \nonumber \\ & =\sum_{u=1}^N \boldsymbol{G}_{n,u}\boldsymbol{x}_{u}+ \boldsymbol{z}_{n}, ~ \forall n.
\end{align}In the above, $\boldsymbol{V}_n = [\boldsymbol{v}(\theta_{n,1}),\ldots,\boldsymbol{v}(\theta_{n,K})] \in \mathbb{C}^{M_t \times K}$ and $\boldsymbol{A}_n = [\boldsymbol{a}(\theta_{n,1}),\ldots,\boldsymbol{a}(\theta_{n,K})] \in \mathbb{C}^{M_r \times K}$ denote the collection of the transmit and receive steering vectors of BS $n$, $\boldsymbol{B}_{n,u} = \textrm{diag}(\boldsymbol{b}_{n,u}) \in \mathbb{C}^{K \times K}$, where $\boldsymbol{b}_{n,u}=[b_{n,u,1},\ldots,b_{n,u,K}]^T \in \mathbb{C}^{K \times 1}$ with $b_{n,u,k}$ being the unknown but deterministic attenuation coefficient from the transmit antenna array of BS $u$ to target $k$ to the receive antenna array of BS $n$ containing both the round-trip path loss and the radar cross-section (RCS) of the target, $\boldsymbol{z}_n \sim \mathcal{CN}(0,\sigma^2 \boldsymbol{I})$ is the Gaussian noise at BS $n$, and
\begin{align}\label{eqn:channel}
    \boldsymbol{G}_{n,u} \triangleq \boldsymbol{A}_{n}\boldsymbol{B}_{n,u} \boldsymbol{V}_{u}^T, ~ \forall n, u.
\end{align}

Due to the limited fronthaul capacity, each BS $n$ needs to compress the echo signal $\boldsymbol{y}_n$ and send the quantization bits to the CU over the fronthaul link, $\forall n$.
In this paper, we consider the Gaussian test channel to model the compression process \cite{Thomas_2006}. Specifically, the quantized signal of BS $n$ is given as
\begin{align}\label{eqn:compression}
    \tilde{\boldsymbol{y}}_{n} & = \boldsymbol{y}_n + \boldsymbol{e}_n  = \sum_{u=1}^{N}  \boldsymbol{G}_{n,u}\boldsymbol{x}_{u} + 
    \boldsymbol{z}_n + \boldsymbol{e}_n, \forall n,
\end{align}
where $\boldsymbol{e}_{n}  \in \mathbb{C}^{M_r \times 1}$ is the Gaussian quantization error vector with zero mean and covariance matrix $\boldsymbol{Q}_{n} \in \mathbb{C}^{M_r \times M_r}$, i.e., $\boldsymbol{e}_n \sim \mathcal{CN}(\boldsymbol{0},\boldsymbol{Q}_n)$.
Based on (\ref{eqn:compression}), the rate (in terms of bits per second per sample) for BS $n$ to transmit $ \tilde{\boldsymbol{y}}_{n}$ to the CU via the fronthaul link is
\begin{align}\label{equ:cons_fronthaul}
    & D_n(\{\boldsymbol{R}_n\}_{n=1}^N,\boldsymbol{Q}_n) =  \nonumber \\ & \mathbb{E}_{\boldsymbol{q}} \left[\log_2\left(\frac{\left|\sum\limits_{u=1}^N\boldsymbol{G}_{n,u}\boldsymbol{R}_u\boldsymbol{G}_{n,u}^H + \sigma^2\boldsymbol{I}+\boldsymbol{Q}_n\right|}{|\boldsymbol{Q}_n|}\right)\right], ~\forall n,
\end{align}where the expectation is over the target location vector $\boldsymbol{q}$, which determines the channel $\boldsymbol{G}_{n,u}$'s.

After receiving the quantized signals from all the $N$ BSs, the CU aims to localize all the $K$ targets. In the following, we first characterize the localization performance under the above networked sensing system as a function of the transmit covariance matrices $\boldsymbol{R}_1,\ldots,\boldsymbol{R}_N$ and the compression noise covariance matrices $\boldsymbol{Q}_1,\ldots,\boldsymbol{Q}_N$, and then optimize the transmit and compression noise covariance matrices to improve the sensing performance subject to the fronthaul capacity constraints.

\section{Posterior Cramér-Rao Bound for Localization}

In this work, we utilize the PCRB as the performance metric to evaluate the performance of the networked sensing system described in the above \cite{Trees_1968}. Specifically, 
let $\boldsymbol{b}^{\rm R} = [(\boldsymbol{b}^{\rm R}_{1,1})^T,\ldots,(\boldsymbol{b}^{\rm R}_{1,N})^T,\ldots,(\boldsymbol{b}^{\rm R}_{N,1})^T,\ldots,(\boldsymbol{b}^{\rm R}_{N,N})^T]^T$ and $\boldsymbol{b}^{\rm I} = [(\boldsymbol{b}^{\rm I}_{1,1})^T,\ldots,(\boldsymbol{b}^{\rm I}_{1,N})^T,\ldots,(\boldsymbol{b}^{\rm I}_{N,1})^T,\ldots,(\boldsymbol{b}^{\rm I}_{N,N})^T]^T$ denote the collection of the real parts and that of the imaginary parts of the attenuation coefficients, respectively, where $\boldsymbol{b}^{\rm R}_{n,u}=[b_{n,,1}^{\rm R},\ldots,b_{n,u,K}^{\rm R}]^T$ and $\boldsymbol{b}^{\rm I}_{n,u}=[b_{n,u,1}^{\rm I},\ldots,b_{n,u,K}^{\rm I}]^T$ with $b_{n,u,k}^{\rm R}$ and $b_{n,u,k}^{\rm I}$ being the real part and the imaginary part of $b_{n,u,k}$, $\forall n,u,k$. Moreover, define $\boldsymbol{\xi} = [\boldsymbol{q}^T,(\boldsymbol{b}^{\rm R})^T,(\boldsymbol{b}^{\rm I})^T]^T$ as the collection of all the unknown parameters to be estimated, and $\tilde{\boldsymbol{y}}=[\tilde{\boldsymbol{y}}_1^T,\ldots,\tilde{\boldsymbol{y}}_N^T]^T$ as the collection of the de-quantized signals at the CU, where $\tilde{\boldsymbol{y}}_n$ is given in (\ref{eqn:compression}). In the rest of this section, we aim to characterize the PCRB for estimating $\boldsymbol{\xi}$ based on the de-quantized signals at the CU $\tilde{\boldsymbol{y}}$ and the prior knowledge of PDF $p_{\boldsymbol{q}}(\boldsymbol{q})$.

In order to derive the PCRB for $\boldsymbol{\xi}$, we first calculate the posterior Fisher information matrix (PFIM) for estimating $\boldsymbol{\xi}$
\begin{align}\label{equ:BFIM_full}
    \boldsymbol{F}_{\boldsymbol{\xi}}(\{\boldsymbol{R}_n,\boldsymbol{Q}_n\}_{n=1}^N) = \boldsymbol{F}_{0,\boldsymbol{\xi}} + \boldsymbol{F}_p,
\end{align}where $\boldsymbol{F}_{0,\boldsymbol{\xi}}$ denotes the FIM corresponding to the de-quantized signals given in (\ref{eqn:compression})
\begin{align}\label{eqn:sig}
    \boldsymbol{F}_{0,\boldsymbol{\xi}} = \mathbb{E}_{\tilde{\boldsymbol{y}},\boldsymbol{q}}\left[ \frac{\partial \log p(\tilde{\boldsymbol{y}}|\boldsymbol{\xi})}{\partial \boldsymbol{\xi}} \Bigg(\frac{\partial \log p(\tilde{\boldsymbol{y}}|\boldsymbol{\xi})}{\partial \boldsymbol{\xi}}\Bigg)^T \right],
\end{align}with
\begin{align}\label{eqn:conditional pdf}
p(\tilde{\boldsymbol{y}}|\boldsymbol{\xi})
 =&~ \prod_{n=1}^{N} \frac{1}{\pi^{M_r}|\sigma^2\boldsymbol{I} + \boldsymbol{Q}_n|} \exp\Bigg[-\Bigg(\tilde{\boldsymbol{y}}_n-\sum_{u=1}^{N}\boldsymbol{G}_{n,u}\boldsymbol{x}_u\Bigg)^H \notag \\
  &\times (\sigma^2\boldsymbol{I} + \boldsymbol{Q}_n)^{-1} \Bigg(\tilde{\boldsymbol{y}}_n-\sum_{u=1}^{N}\boldsymbol{G}_{n,u}\boldsymbol{x}_u\Bigg)\Bigg],
\end{align}and $\boldsymbol{F}_p$ denotes the FIM corresponding to the prior knowledge about location distribution
\begin{align}\label{eqn:dis}
\boldsymbol{F}_p & = \mathbb{E}_{\boldsymbol{q}}\left[ \frac{\partial \log p_{\boldsymbol{q}}(\boldsymbol{q})}{\partial \boldsymbol{\xi}} \Bigg(\frac{\partial \log p_{\boldsymbol{q}}(\boldsymbol{q})}{\partial \boldsymbol{\xi}}\Bigg)^T \right] \nonumber \\
& = \begin{bmatrix}
        \mathbb{E}_{\boldsymbol{q}}\Bigg\{\frac{\partial \log p_{\boldsymbol{q}}(\boldsymbol{q})}{\partial \boldsymbol{q}}\Big(\frac{\partial \log p_{\boldsymbol{q}}(\boldsymbol{q})}{\partial \boldsymbol{q}}\Big)^T\Bigg\} & \boldsymbol{0} \\
        \boldsymbol{0} & \boldsymbol{0}
    \end{bmatrix}.
\end{align}

Given any PDF $p_{\boldsymbol{q}}(\boldsymbol{q})$, $\boldsymbol{F}_p$ can be computed based on (\ref{eqn:dis}). Therefore, the remaining job to characterize $\boldsymbol{F}_{\boldsymbol{\xi}}(\{\boldsymbol{R}_n,\boldsymbol{Q}_n\}_{n=1}^N)$ is to quantify $\boldsymbol{F}_{0,\boldsymbol{\xi}}$ given in (\ref{eqn:sig}). It is observed from (\ref{eqn:channel}) that $\boldsymbol{G}_{n,u}$'s are functions of $\boldsymbol{\theta} = [\boldsymbol{\theta}_1^T,\dots,\boldsymbol{\theta}_N^T]^{T}$ with $ \boldsymbol{\theta}_n =[\theta_{n,1},\ldots,\theta_{n,K}]^T$. Moreover, because $\boldsymbol{\theta}$ is a complicated function of the target locations $\boldsymbol{q}$ as shown in (\ref{equ:theta_xy}), $p(\tilde{\boldsymbol{y}}|\boldsymbol{\xi})$ is a complicated function of $\boldsymbol{q}$, and it is challenging to calculate its derivative over $\boldsymbol{q}$. We apply the chain rule to tackle this challenge. Specifically, define $\boldsymbol{\zeta} = [\boldsymbol{\theta}^T, (\boldsymbol{b}^{\rm R})^T, (\boldsymbol{b}^{\rm I})^T]^T$. 
Note that different from parameter vector $\boldsymbol{\xi}$, target location $\boldsymbol{q}$ has been changed to target angle $\boldsymbol{\theta}$ in the parameter vector $\boldsymbol{\zeta}$. 
Moreover, define
\begin{align}
\boldsymbol{u}_{n,k} = \frac{\partial \theta_{n,k}}{\partial \boldsymbol{q}_k} = \left[\frac{\partial \theta_{n,k}}{\partial q_k^x}, \frac{\partial \theta_{n,k}}{\partial q_k^y}\right]^T, ~ \forall n, k,
\end{align}where according to (\ref{equ:theta_xy}),
\begin{align}
\frac{\partial \theta_{n,k}}{\partial q_n^x} &= \frac{1}{(1+\tan^2\theta_{n,k})(q_k^y-\rho_n^y)}, \\
\frac{\partial \theta_{n,k}}{\partial q_n^y} &= -\frac{1}{(1+\cot^2\theta_{n,k})(q_k^x-\rho_n^x)}.
\end{align}Then, based on the chain rule, $\boldsymbol{F}_{0,\boldsymbol{\xi}}$ can be obtained as follows
\begin{align}\label{equ:F_0xi}
    \boldsymbol{F}_{0,\boldsymbol{\xi}} =  \mathbb{E}_{\boldsymbol{\xi}}[\boldsymbol{U}\boldsymbol{F}_{0,\boldsymbol{\zeta}}\boldsymbol{U}^T].
\end{align}In the above, $\boldsymbol{U}$ is the Jacobian matrix given by
\begin{align}\label{equ:trans_mat}
    \boldsymbol{U} \triangleq \frac{\partial \boldsymbol{\zeta}}{\partial \boldsymbol{\xi}} &= \begin{bmatrix}
        \hat{\boldsymbol{U}} & \boldsymbol{0} \\
        \boldsymbol{0} & \boldsymbol{I}
    \end{bmatrix} \in \mathbb{C}^{(2K+N^2K) \times (NK+N^2K)},
\end{align}where
\begin{align}
    \hat{\boldsymbol{U}} &= [\boldsymbol{U}_1,\dots,\boldsymbol{U}_N] \in \mathbb{C}^{2K \times NK}, \\
    \boldsymbol{U}_n &= \textbf{blkdiag}\left\{ \boldsymbol{u}_{n,1},\dots,\boldsymbol{u}_{n,K}\right\} \notag \\
    &= \begin{bmatrix}
        \boldsymbol{u}_{n,1} & \dots & \boldsymbol{0} \\
        \vdots & \ddots & \vdots \\
        \boldsymbol{0} & \dots & \boldsymbol{u}_{n,K}
       \end{bmatrix} \in \mathbb{C}^{2K \times K}, \forall n.
\end{align}
Moreover, $\boldsymbol{F}_{0,\boldsymbol{\zeta}}$ is defined as
\begin{align}\label{eqn:zeta1}
\boldsymbol{F}_{0,\boldsymbol{\zeta}}=\mathbb{E}_{\tilde{\boldsymbol{y}}} \left[ \frac{\partial \log p(\tilde{\boldsymbol{y}}|\boldsymbol{\zeta})}{\partial \boldsymbol{\zeta}} \Bigg(\frac{\partial \log p(\tilde{\boldsymbol{y}}|\boldsymbol{\zeta})}{\partial \boldsymbol{\zeta}}\Bigg)^T \right],
\end{align}where $p(\tilde{\boldsymbol{y}}|\boldsymbol{\zeta})$ has the similar form to $p(\tilde{\boldsymbol{y}}|\boldsymbol{\xi})$, but we should treat $\boldsymbol{G}_{n,u}$'s given in (\ref{eqn:channel}) as functions of $\boldsymbol{\theta}$ instead of $\boldsymbol{q}$.

The explicit expression for $\boldsymbol{F}_{0,\boldsymbol{\zeta}}$ can be given by the following proposition.
\begin{proposition}\label{prop:FIM_MC}
Define $\boldsymbol{F}_1$, $\boldsymbol{F}_2$, and $\boldsymbol{F}_3$ as (\ref{equ:F1}) -- (\ref{equ:F3}) on the top of the next page, where
\begin{figure*}
\begin{small}
\begin{align}
    \boldsymbol{F}_1 &= \begin{bmatrix}
                         \boldsymbol{F}_1[1,1] & \dots & \boldsymbol{F}_1[1,N] \\
                         \vdots & \ddots & \vdots \\
                         \boldsymbol{F}_1[N,1] & \dots & \boldsymbol{F}_1[N,N]
                       \end{bmatrix} \in \mathbb{C}^{NK \times NK}, \label{equ:F1} \\
    \boldsymbol{F}_2 &= \begin{bmatrix}
                         \boldsymbol{F}_2[1,1,1] & \dots  & \boldsymbol{F}_2[1,1,N] & \dots & \boldsymbol{F}_2[1,N,1] & \dots  & \boldsymbol{F}_2[1,N,N] \\
                         \vdots                  & \ddots & \vdots                  & \dots & \vdots                  & \ddots & \vdots                  \\
                         \boldsymbol{F}_2[N,1,1] & \dots  & \boldsymbol{F}_2[N,1,N] & \dots & \boldsymbol{F}_2[N,N,1] & \dots  & \boldsymbol{F}_2[N,N,N]
                       \end{bmatrix} \in \mathbb{C}^{NK \times N^2K}, \label{equ:F2} \\
    \boldsymbol{F}_3 &= \begin{bmatrix}
                         \boldsymbol{F}_3[1,1,1,1] & \dots & \boldsymbol{F}_3[1,1,1,N] & \dots & \boldsymbol{F}_3[1,1,N,1] & \dots & \boldsymbol{F}_3[1,1,N,N] \\
                         \vdots                    & \ddots & \vdots                   & \ddots & \vdots                    & \ddots & \vdots \\
                         \boldsymbol{F}_3[1,N,1,1] & \dots & \boldsymbol{F}_3[1,N,1,N] & \dots & \boldsymbol{F}_3[1,N,N,1] & \dots & \boldsymbol{F}_3[1,N,N,N] \\
                         \vdots                    & \ddots & \vdots                   & \ddots & \vdots                    & \ddots & \vdots \\
                         \boldsymbol{F}_3[N,1,1,1] & \dots & \boldsymbol{F}_3[N,1,1,N] & \dots & \boldsymbol{F}_3[N,1,N,1] & \dots & \boldsymbol{F}_3[N,1,N,N] \\
                         \vdots                    & \ddots & \vdots                   & \ddots & \vdots                    & \ddots & \vdots \\
                         \boldsymbol{F}_3[N,N,1,1] & \dots & \boldsymbol{F}_3[N,N,1,N] & \dots & \boldsymbol{F}_3[N,N,N,1] & \dots & \boldsymbol{F}_3[N,N,N,N]
                       \end{bmatrix} \in \mathbb{C}^{N^2K \times N^2K}. \label{equ:F3}
\end{align}
\end{small}
\hrule
\end{figure*}
\begin{small}
\begin{align}
    &\boldsymbol{F}_{1}[i,i] = \sum_{u=1}^{N} \Big[\Big(\boldsymbol{A}_u^H\boldsymbol{O}^{-1}_u\boldsymbol{A}_u) \odot (\boldsymbol{B}^{*}_{u,i}\dot{\boldsymbol{V}}_{i}^H\boldsymbol{R}_{i}^*\dot{\boldsymbol{V}}_{i}\boldsymbol{B}_{u,i}\Big) \Big] \notag \\
    &+ \sum_{u=1}^{N} \Big[\Big(\dot{\boldsymbol{A}}_i^H\boldsymbol{O}^{-1}_i\dot{\boldsymbol{A}}_i) \odot (\boldsymbol{B}_{i,u}^{*}\boldsymbol{V}_{u}^{H}\boldsymbol{R}_{u}^*\boldsymbol{V}_{u}\boldsymbol{B}_{i,u}\Big)\Big] \notag \\
    &+ (\boldsymbol{A}_i^H\boldsymbol{O}^{-1}_i\dot{\boldsymbol{A}}_i) \odot (\boldsymbol{B}^{*}_{i,i}\dot{\boldsymbol{V}}_{i}^H\boldsymbol{R}_{i}^*\boldsymbol{V}_{i}\boldsymbol{B}_{i,i}) \notag \\
    &+ (\dot{\boldsymbol{A}}_i^H\boldsymbol{O}^{-1}_i\boldsymbol{A}_i) \odot (\boldsymbol{B}^{*}_{i,i}\boldsymbol{V}_{i}^H\boldsymbol{R}_{i}^*\dot{\boldsymbol{V}}_{i}\boldsymbol{B}_{i,i}), ~\forall i, \label{equ:F1_ii} \\
    &\boldsymbol{F}_{1}[i,n] = (\dot{\boldsymbol{A}}_i^H\boldsymbol{O}^{-1}_i\boldsymbol{A}_i) \odot (\boldsymbol{B}^{*}_{i,n}\boldsymbol{V}_{n}^H\boldsymbol{R}_{n}^*\dot{\boldsymbol{V}}_{n}\boldsymbol{B}_{i,n}) \notag \\
    &+ (\boldsymbol{A}_n^H\boldsymbol{O}^{-1}_n\dot{\boldsymbol{A}}_n) \odot (\boldsymbol{B}^{*}_{n,i}\dot{\boldsymbol{V}}_{i}^H\boldsymbol{R}_{i}^*\boldsymbol{V}_{i}\boldsymbol{B}_{n,i}) , ~\forall i \ne n, \label{equ:F1_in} \\
    &\boldsymbol{F}_{2}[i,i,i] = (\dot{\boldsymbol{A}}_i^H\boldsymbol{O}^{-1}_i\boldsymbol{A}_i) \odot (\boldsymbol{B}_{i,i}^{*}\boldsymbol{V}_{i}^H\boldsymbol{R}_{i}^*\boldsymbol{V}_{i}) \notag \\
    &+ (\boldsymbol{A}_i^H\boldsymbol{O}^{-1}_i\boldsymbol{A}_i) \odot (\boldsymbol{B}_{i,i}^{*}\dot{\boldsymbol{V}}_{i}^H\boldsymbol{R}_{i}^*\boldsymbol{V}_{i}), ~\forall i, \label{equ:F2_iii} \\
    &\boldsymbol{F}_{2}[i,i,n] = (\dot{\boldsymbol{A}}_i^H\boldsymbol{O}^{-1}_i\boldsymbol{A}_i) \odot (\boldsymbol{B}_{i,n}^{*}\boldsymbol{V}_{n}^H\boldsymbol{R}_{n}^*\boldsymbol{V}_{n}), ~\forall i \ne n, \label{equ:F2_iin} \\
    &\boldsymbol{F}_{2}[i,n,i] = (\boldsymbol{A}_n^H\boldsymbol{O}^{-1}_n\boldsymbol{A}_n) \odot (\boldsymbol{B}_{n,i}^{*}\dot{\boldsymbol{V}}_{i}^{H}\boldsymbol{R}_{i}^{*}\boldsymbol{V}_{i}), ~\forall i \ne n, \label{equ:F2_ini} \\
    &\boldsymbol{F}_{2}[i,n,k] = \mathbf{0}, ~\forall i \ne n, i \ne k, \label{equ:F2_ink} \\
    &\boldsymbol{F}_{3}[n,i,n,i] = (\boldsymbol{A}_{n}^H\boldsymbol{O}^{-1}_n\boldsymbol{A}_{n}) \odot (\boldsymbol{V}_{i}^{H}\boldsymbol{R}_{i}^*\boldsymbol{V}_{i}), ~\forall i, \label{equ:F3_nini} \\
    &\boldsymbol{F}_{3}[n,i,n,k] = \mathbf{0}, ~\forall i \ne k, \label{equ:F3_nink} \\
    &\boldsymbol{F}_{3}[n,i,u,k] = \mathbf{0}, ~\forall n \ne u, \label{equ:F3_niuk}
\end{align}
\end{small}with $\dot{\boldsymbol{A}}_n = \left[\frac{\partial \boldsymbol{a}(\theta_{n,1})}{\partial \theta_{n,1}},\ldots,\frac{\partial \boldsymbol{a}(\theta_{n,K})}{\partial \theta_{n,K}}\right] \in \mathbb{C}^{M_r \times K}$, $\dot{\boldsymbol{V}}_n = \left[\frac{\partial \boldsymbol{v}(\theta_{n,1})}{\partial \theta_{n,1}},\ldots,\frac{\partial \boldsymbol{v}(\theta_{n,K})}{\partial \theta_{n,K}}\right] \in \mathbb{C}^{M_t \times K}$, and $\boldsymbol{O}_n = \sigma^2\boldsymbol{I} + \boldsymbol{Q}_n \in \mathbb{C}^{M_r \times M_r}$, $\forall n$.
Then, $\boldsymbol{F}_{0,\boldsymbol{\zeta}}$ given in (\ref{eqn:zeta1}) can be expressed as
\begin{align}\label{equ:F_theta_rcs}
    \boldsymbol{F}_{0,\boldsymbol{\zeta}} = 2\begin{bmatrix}
    \Re\{\boldsymbol{F}_1\} & \Re\{\boldsymbol{F}_2\} & -\Im\{\boldsymbol{F}_2\} \\
    \Re\{\boldsymbol{F}_2\}^T & \Re\{\boldsymbol{F}_3\} & -\Im\{\boldsymbol{F}_3\} \\
    -\Im\{\boldsymbol{F}_2\}^T & -\Im\{\boldsymbol{F}_3\}^T & \Re\{\boldsymbol{F}_3\}
    \end{bmatrix},
\end{align}where given each matrix $\boldsymbol{X}$, $\Re\{\boldsymbol{X}\}$ and $\Im\{\boldsymbol{X}\}$ are the matrices with each element being the real part and imaginary part of the element in $\boldsymbol{X}$, respectively.
\end{proposition}
\begin{IEEEproof}
    Please refer to Appendix \ref{app:proA}.
\end{IEEEproof}


After characterizing the PFIM $\boldsymbol{F}_{\boldsymbol{\xi}}(\{\boldsymbol{R}_n,\boldsymbol{Q}_n\}_{n=1}^N)$ shown in (\ref{equ:BFIM_full}), where $\boldsymbol{F}_{0,\boldsymbol{\xi}}$ is obtained via Proposition \ref{prop:FIM_MC} and $\boldsymbol{F}_p$ is obtained via (\ref{eqn:dis}), the PCRB for the mean-squared error (MSE) of any estimator of $\boldsymbol{q}$, denoted by $\hat{\boldsymbol{q}}$, can be expressed as
\begin{align}\label{equ:PCRB_q}
\mathbb{E}_{\boldsymbol{q}}[\|\hat{\boldsymbol{q}}-\boldsymbol{q}\|^2]\geq \sum_{i=1}^{2K} [(\boldsymbol{F}_{\boldsymbol{\xi}}(\{\boldsymbol{R}_n,\boldsymbol{Q}_n\}_{n=1}^N))^{-1}](i,i).
\end{align}

\begin{remark}
We want to emphasize that our result is a generalization of the result in \cite{Li_2008_TSP}. 
Specifically, the work \cite{Li_2008_TSP} has characterized the CRB for estimating the unknown but deterministic locations of targets for the special case of one BS, i.e., $N=1$, while our result holds for estimating the random target locations in the general case of networked sensing via cooperation among multiple BSs.  
\end{remark}


\section{Optimizing Transmission and Compression Strategies for PCRB Minimization}
In this section, we aim to design the transmit covariance matrices and the compression noise covariance matrices of all the BSs, i.e., $\boldsymbol{R}_n$ and $\boldsymbol{Q}_n$, $\forall n$, to minimize the PCRB for localization, subject to the individual transmit power constraint and fronthaul capacity constraint of each BS. Specifically, the above problem can be formulated as
\begin{subequations}
\begin{align}
    (\text{P1})\min_{\{\boldsymbol{R}_n,\boldsymbol{Q}_n\}_{n=1}^N} &~ \sum_{i=1}^{2K} \left[\left(\boldsymbol{F}_{\boldsymbol{\xi}}\left(\{\boldsymbol{R}_n,\boldsymbol{Q}_n\}_{n=1}^N\right)\right)^{-1}\right](i,i) \\
    \text{s.t.}~~~~~ &~ \text{tr}(\boldsymbol{R}_n) \le \bar{P}_n, ~ \forall n, \label{eqn:1}\\
    \quad &~ D_n(\{\boldsymbol{R}_n\}_{n=1}^N,\boldsymbol{Q}_n)\leq \bar{D}_n, ~ \forall n, \label{eqn:2} \\ 
    \quad &~ \boldsymbol{Q}_{n} \succeq \boldsymbol{0}, ~ \forall n, \label{eqn:3}
\end{align}
\end{subequations}where $\bar{P}_n$ is transmit power constraint for BS $n$, $\bar{D}_n$ is the fronthaul capacity constraint for BS $n$, and $D_n(\{\boldsymbol{R}_n\}_{n=1}^N,\boldsymbol{Q}_n)$ is the fronthaul transmission rate of BS $n$ given in (\ref{equ:cons_fronthaul}), $\forall n$.

Our aim is to design the BSs' transmit covariance matrices, i.e., $\boldsymbol{R}_n$'s, and compression noise covariance matrices, i.e., $\boldsymbol{Q}_n$'s, in Problem (P1). Note that the objective function of Problem (P1) is a complicated function of $\boldsymbol{R}_n$'s and $\boldsymbol{Q}_n$'s. To tackle this challenge, we adopt the Schur complement technique to equivalently transform Problem (P1) to the following problem
\begin{subequations}
\begin{align}
    (\text{P2})\min_{\{\boldsymbol{R}_n,\boldsymbol{Q}_n\}_{n=1}^N,\{t_i\}_{i=1}^{2K}} &~ \sum_{i=1}^{2K} t_i \\
    \text{s.t.}~~~~~~~~~ &~ \begin{bmatrix}
        \boldsymbol{F}_{\boldsymbol{\xi}}(\{\boldsymbol{R}_n,\boldsymbol{Q}_n\}_{n=1}^N) & \boldsymbol{e}_i \\
        \boldsymbol{e}_i^T & t_i
    \end{bmatrix} \succeq \boldsymbol{0}, ~ \forall i, \label{equ:cons_F_CRB_P2} \\
    \quad &~ {\rm (\ref{eqn:1})-(\ref{eqn:3})}. \nonumber
\end{align}
\end{subequations}
where $t_1,\ldots,t_{2K}$ are auxiliary variables and $\boldsymbol{e}_i$ is the $i$th column of the identity matrix $\boldsymbol{I}$, $\forall i$. Problem (P2) is still challenging, because all the variables are coupled together. In this work, we design an AO-based algorithm to iteratively solve Problem (P2). Specifically, at each iteration of the AO-based algorithm, we first fix $\boldsymbol{Q}_n$'s and optimize $\boldsymbol{R}_n$'s and $t_i$'s in Problem (P2), and then fix $\boldsymbol{R}_n$'s and optimize $\boldsymbol{Q}_n$'s and $t_i$'s in Problem (P2). This can guarantee that the objective value of Problem (P2) will be decreased after each iteration. The algorithm will terminate at some iteration when the decrease in the objective value is smaller than some threshold. In the following, we respectively show how to optimize $\boldsymbol{R}_n$'s and $t_i$'s given $\boldsymbol{Q}_n$'s and how to optimize $\boldsymbol{Q}_n$'s and $t_i$'s given $\boldsymbol{R}_n$'s in Problem (P2).

\subsection{Optimization of Transmission Covariance Matrices}\label{sec:opt_Rn}
Given any $\boldsymbol{Q}_n=\bar{\boldsymbol{Q}}_n$'s, we aim to solve the following sub-problem of Problem (P2)
\begin{subequations}\label{opt:P2.1}
\begin{align}
    (\text{P2.1})\min_{\{\boldsymbol{R}_n\}_{n=1}^N,\{t_i\}_{i=1}^{2K}} &~ \sum_{i=1}^{2K} t_i \\
    \text{s.t.}~~~~~~~ &~ D_n(\{\boldsymbol{R}_n\}_{n=1}^N,\bar{\boldsymbol{Q}}_n) \leq \bar{D}_n, ~ \forall n, \label{eqn:FC_P2.1} \\
    \quad &~{\rm (\ref{equ:cons_F_CRB_P2}), ~ (\ref{eqn:1})}. \nonumber 
\end{align}
\end{subequations}
Problem (P2.1) is a non-convex problem because $D_n(\{\boldsymbol{R}_n\}_{n=1}^N,\bar{\boldsymbol{Q}}_n)$'s given in (\ref{equ:cons_fronthaul}) are concave functions, instead of convex functions, of $\boldsymbol{R}_n$'s. We apply the SCA technique to tackle this challenge, whose idea is to successively optimize the convex approximation of the original problem. Specifically, because $\hat{D}_n (\{\boldsymbol{R}_n\}_{n=1}^N,\bar{\boldsymbol{Q}}_n)$ is a concave function over the transmit covariance matrices $\{\boldsymbol{R}_n\}_{n=1}^N$, given any $\tilde{\boldsymbol{R}}_n \succeq 0, \forall n$, the following inequality always holds

\begin{align}\label{equ:Dn_ieq_R}
& D_n(\{\boldsymbol{R}_n\}_{n=1}^N,\bar{\boldsymbol{Q}}_n) \leq \hat{D}_n (\{\boldsymbol{R}_n,\tilde{\boldsymbol{R}}_n\}_{n=1}^N,\bar{\boldsymbol{Q}}_n), ~ \forall n,
\end{align}where
\begin{align}
    & \hat{D}_n (\{\boldsymbol{R}_n,\tilde{\boldsymbol{R}}_n\}_{n=1}^N,\bar{\boldsymbol{Q}}_n) =  \nonumber \\ 
    &\frac{1}{\log2}\mathbb{E}_{\boldsymbol{q}}\Bigg\{\text{tr}\Bigg[ \bar{\boldsymbol{\Sigma}}_n^{-1}\Bigg(\sum\limits_{u=1}^N\boldsymbol{G}_{n,u}\boldsymbol{R}_u\boldsymbol{G}_{n,u}^H + \sigma^2\boldsymbol{I}+\bar{\boldsymbol{Q}}_n\Bigg) \Bigg]\Bigg\} \notag \\
    &- \frac{M_r}{\log2}  + \mathbb{E}_{\boldsymbol{q}}\Big[\log_2|\bar{\boldsymbol{\Sigma}}_n|\Big] - \log_2|\bar{\boldsymbol{Q}}_n|, \quad \forall n,  \label{equ:relax_FC_optR} \\
    &\bar{\boldsymbol{\Sigma}}_n = \sum\limits_{u=1}^N\boldsymbol{G}_{n,u}\tilde{\boldsymbol{R}}_u\boldsymbol{G}_{n,u}^H + \sigma^2\boldsymbol{I}_n+\bar{\boldsymbol{Q}}_n, \quad \forall n. \label{eqn:Sigma_n}
\end{align}Here, the first term at the right hand side of the equation (\ref{equ:relax_FC_optR}) is a linear function on the transmit covariance matrix $\boldsymbol{R}_n$'s, and the other three terms are all constants.
Based on the above inequality, given any $\tilde{\boldsymbol{R}}_n\succeq \boldsymbol{0}$, $\forall n$, we aim to solve the following relaxed problem of Problem (P2.1)
\begin{subequations}\label{opt:relax_P2.1}
\begin{align}
\min_{\{\boldsymbol{R}_n\}_{n=1}^N,\{t_i\}_{i=1}^{2K}}~ & \sum_{i=1}^{2K} t_i \label{eqn:P2} \\
    \text{s.t.}~~~~~~~ & \hat{D}_n (\{\boldsymbol{R}_n,\tilde{\boldsymbol{R}}_n\}_{n=1}^N,\bar{\boldsymbol{Q}}_n)\leq \bar{D}_n, \forall n, \label{eqn:4} \\
    & {\rm (\ref{eqn:1}), ~ (\ref{equ:cons_F_CRB_P2})}. \nonumber
\end{align}
\end{subequations}
Note that in Problem (\ref{opt:relax_P2.1}), the non-convex constraint (\ref{eqn:2}) in Problem (P2.1) is replaced by the convex constraint (\ref{eqn:4}). 
Because Problem (\ref{opt:relax_P2.1}) is a convex problem, we can utilize the interior-point method to globally solve it. Becauses of (\ref{equ:Dn_ieq_R}), the solution to Problem (\ref{opt:relax_P2.1}) is a feasible solution to Problem (P2.1).

We summarize the SCA-based algorithm to solve Problem (P2.1) via Problem (\ref{opt:relax_P2.1}) in Algorithm \ref{table1}. According to the theory about the SCA technique, Algorithm \ref{table1} can yield a locally optimal solution to Problem (P2.1).

\begin{table}[htp]
\begin{center}
\caption{\textbf{Algorithm I}: SCA-Based Algorithm for Problem (P2.1)} \vspace{-0.2cm}
 \hrule
\vspace{0.2cm}
\begin{itemize}
\item[1.] Initialize: Set $\boldsymbol{R}^{(0)}_n$ as the initial matrix of $\boldsymbol{R}_n$, $\forall n$, and $m=0$, where $m$ denotes the index of iteration;
\item[2.] Repeat
\begin{itemize}
\item[a.] Set $m = m + 1$;
\item[b.] Update $\tilde{\boldsymbol{R}}_n=\boldsymbol{R}^{(m-1)}_n$, $\forall n$, in Problem (\ref{opt:relax_P2.1});
\item[c.] Update $\boldsymbol{R}^{(m)}_1,\ldots,\boldsymbol{R}^{(m)}_N$ as the optimal solution of Problem (\ref{opt:relax_P2.1});
\end{itemize}
\item[3.] Until convergence.
\end{itemize}
\vspace{0.2cm} \hrule \label{table1}
\end{center}
\end{table}

\subsection{Optimization of Compression Noise Covariance Matrices}\label{sec:opt_On}
Next, we show how to optimize the compression noise covariance matrices. Given any $\boldsymbol{R}_n=\bar{\boldsymbol{R}}_n$'s, we aim to solve the following sub-problem of Problem (P2)
\begin{subequations}
\begin{align}
    (\text{P2.2})\min_{\{\boldsymbol{Q}_n\}_{n=1}^N,\{t_i\}_{i=1}^{2K}} &~ \sum_{i=1}^{2K} t_i \\
    \text{s.t.}~~~~~ &~ {\rm (\ref{equ:cons_F_CRB}), ~ (\ref{eqn:2}), ~ (\ref{eqn:3})}. \nonumber
\end{align}
\end{subequations}Problem (P2.2) is non-convex because $\boldsymbol{F}_{\boldsymbol{\xi}}(\{\bar{\boldsymbol{R}}_n,\boldsymbol{Q}_n\}_{n=1}^N)$ is not a concave function of $\boldsymbol{Q}_n$'s in (\ref{equ:cons_F_CRB}), and $D_n(\{\bar{\boldsymbol{R}}_n\}_{n=1}^N,\boldsymbol{Q}_n)$ is not a convex function of $\boldsymbol{Q}_n$ in (\ref{eqn:2}). To tackle the above challenge, we first change the optimization variables. Specifically, define
\begin{align}\label{eqn:T}
\boldsymbol{T}_n=(\sigma^2\boldsymbol{I}+\boldsymbol{Q}_n)^{-1}, ~ \forall n.
\end{align}
By replacing $\boldsymbol{Q}_n$ with $\boldsymbol{T}_n^{-1}-\sigma^2\boldsymbol{I}$, $\forall n$, in $\boldsymbol{F}_{\boldsymbol{\xi}}(\{\boldsymbol{R}_n,\boldsymbol{Q}_n\}_{n=1}^N)$ that is characterized in Proposition \ref{prop:FIM_MC}, we can get  $\boldsymbol{F}_{\boldsymbol{\xi}}(\{\boldsymbol{R}_n,\boldsymbol{T}_n\}_{n=1}^N)$. Moreover, the fronthaul compression rate given in (\ref{equ:cons_fronthaul}) becomes
\begin{align}
 D_n(\{\boldsymbol{R}_n\}_{n=1}^N,\boldsymbol{T}_n) =& \mathbb{E}_{\boldsymbol{q}}\left[\log_2\Bigg|\sum\limits_{u=1}^N\boldsymbol{G}_{n,u}\boldsymbol{R}_u\boldsymbol{G}_{n,u}^H + \boldsymbol{T}_n^{-1} \Bigg|\right] \nonumber\\
& -\log_2\Big|\boldsymbol{T}_n^{-1}-\sigma^2\boldsymbol{I}\Big|, ~ \forall n.
\end{align}Therefore, if we change the design variables from $\boldsymbol{Q}_n$'s to $\boldsymbol{T}_n$'s, then Problem (P2.2) is equivalent to
\begin{subequations}\label{eqn:pro fronthaul}
\begin{align}
    \min_{\{\boldsymbol{T}_n\}_{n=1}^N,\{t_i\}_{i=1}^{2K}} &~ \sum_{i=1}^{2K} t_i \label{eqn:P3} \\
    \text{s.t.}~~~~~ & \begin{bmatrix}
        \boldsymbol{F}_{\boldsymbol{\xi}}(\{\bar{\boldsymbol{R}}_n,\boldsymbol{T}_n\}_{n=1}^N) & \boldsymbol{e}_i \\
        \boldsymbol{e}_i^T & t_i
    \end{bmatrix} \succeq \boldsymbol{0}, ~ \forall i, \label{equ:cons_F_CRB T} \\
    &  D_n(\{\bar{\boldsymbol{R}}_n\}_{n=1}^N,\boldsymbol{T}_n)\leq \bar{D}_n, ~ \forall n, \label{eqn:fronthaul} \\
    &  \sigma^{-2}\boldsymbol{I} - \boldsymbol{T}_n \succeq \boldsymbol{0}, \forall n. \label{eqn:C}
\end{align}
\end{subequations}Now, (\ref{equ:cons_F_CRB T}) is convex over $\boldsymbol{T}_n$'s. However, (\ref{eqn:fronthaul}) is still non-convex because $ D_n(\{\bar{\boldsymbol{R}}_n\}_{n=1}^N,\boldsymbol{T}_n)$ is a concave function, instead of convex function, of $\boldsymbol{T}_n$. Similar to the optimization of $\boldsymbol{R}_n$'s, we can apply the SCA technique to tackle this challenge. Specifically, the relaxed problem of Problem (\ref{eqn:pro fronthaul}) is
\begin{subequations}\label{eqn:relax_P2.2}
\begin{align}
    \min_{\{\boldsymbol{T}_n\}_{n=1}^N,\{t_i\}_{i=1}^{2K}} &~ \sum_{i=1}^{2K} t_i \label{eqn:P4} \\
    \text{s.t.}~~~~~ & \tilde{D}_n(\{\bar{\boldsymbol{R}}_n\}_{n=1}^N,\bar{\boldsymbol{T}}_n,\boldsymbol{T}_n) \leq \bar{D}_n, ~ \forall n, \label{eqn:relax_fronthaul} \\
    & {\rm (\ref{equ:cons_F_CRB T}), ~ (\ref{eqn:C}) }. \nonumber
\end{align}
\end{subequations}
where
\begin{align}
    &~\tilde{D}_n(\{\bar{\boldsymbol{R}}_n\}_{n=1}^N,\bar{\boldsymbol{T}}_n,\boldsymbol{T}_n) = \notag \\
    &\frac{1}{\log2}\mathbb{E}_{\boldsymbol{q}}\Big\{\text{tr}\left[ \bar{\boldsymbol{\Omega}}_n^{-1}\Big( \boldsymbol{I} + \bar{\boldsymbol{J}}_n^{\frac{1}{2}}\boldsymbol{T}_n\bar{\boldsymbol{J}}_n^{\frac{1}{2}} \Big) \right]\Big\} - \frac{M_r}{\log2} \notag \\
    &+ \mathbb{E}_{\boldsymbol{q}}\Big[\log_2|\bar{\boldsymbol{\Omega}}_n|\Big] - \log_2 \left| \boldsymbol{I} - \sigma^2\boldsymbol{T}_n \right|, ~\forall n, \label{equ:relax_FC_optT} \\
    &~\bar{\boldsymbol{J}}_n = \sum\limits_{u=1}^N\boldsymbol{G}_{n,u}\bar{\boldsymbol{R}}_u\boldsymbol{G}_{n,u}^H, ~\forall n, \\
    &~\bar{\boldsymbol{\Omega}}_n = \boldsymbol{I} + \bar{\boldsymbol{J}}_n^{\frac{1}{2}}\bar{\boldsymbol{T}}_n\bar{\boldsymbol{J}}_n^{\frac{1}{2}}, ~\forall n. \label{eqn:Omega_n}
\end{align} Similar to (\ref{equ:relax_FC_optR}), the function $\tilde{D}_n(\{\bar{\boldsymbol{R}}_n\}_{n=1}^N,\bar{\boldsymbol{T}}_n,\boldsymbol{T}_n)$ is also a linear function over $\boldsymbol{T}_n$'s.
Therefore, Problem (\ref{eqn:relax_P2.2}) is a convex problem and can be solved by interior-point method. Then, the SCA-based algorithm to solve Problem (\ref{eqn:pro fronthaul}) is summarized in Algorithm \ref{table2}. After the solution to Problem (\ref{eqn:pro fronthaul}) is obtained, the compression noise covariance matrices can be obtained via (\ref{eqn:T}).
\begin{table}[htp]
\begin{center}
\caption{\textbf{Algorithm II}: SCA-Based Algorithm for Problem (P2.2)} \vspace{-0.2cm}
 \hrule
\vspace{0.2cm}
\begin{itemize}
\item[1.] Initialize: Set $\boldsymbol{T}^{(0)}_n$ as the initial matrix of $\boldsymbol{T}_n$, $\forall n$, and $m=0$, where $m$ denotes the index of iteration;
\item[2.] Repeat
\begin{itemize}
\item[a.] $m = m + 1$;
\item[b.] Update $\bar{\boldsymbol{T}}_n=\boldsymbol{T}^{(m-1)}_n$, $\forall n$, in Problem (\ref{eqn:relax_P2.2});
\item[c.] Update $\boldsymbol{T}^{(m)}_1,\ldots,\boldsymbol{T}^{(m)}_N$ as the optimal solution of Problem (\ref{eqn:relax_P2.2});
\end{itemize}
\item[3.] Until convergence. Set the compression noise covariance matrices as $\boldsymbol{Q}_n=(\boldsymbol{T}^{(m)}_n)^{-1}-\sigma^2 \boldsymbol{I}$, $\forall n$.
\end{itemize}
\vspace{0.2cm} \hrule \label{table2}
\end{center}\vspace{-10pt}
\end{table}

Finally, the alternating optimization based algorithm to solve Problem (P2) is summarized in Algorithm \ref{table3}.
\begin{table}[htp]
\begin{center}
\caption{\textbf{Algorithm III}: Overall Algorithm for Problem (P2)} \vspace{-0.2cm}
 \hrule
\vspace{0.2cm}
\begin{itemize}
\item[1.] Initialize: Set $\boldsymbol{R}_n^{(0)}$ and $\boldsymbol{Q}_n^{(0)}$ as the initial matrices of $\boldsymbol{R}_{n}$ and $\boldsymbol{Q}_{n}$, $\forall n$, and $m=0$, where $m$ denotes the index of iteration;
\item[2.] Repeat
\begin{itemize}
\item[a.] $m = m + 1$;
\item[b.] Set $\bar{\boldsymbol{Q}}_{n} = \boldsymbol{Q}_{n}^{(m-1)}$, $\forall n$, and update $\boldsymbol{R}^{(m)}_{n}$, $\forall n$, as the solution to Problem (P2.1) obtained via Algorithm \ref{table1};
\item[c.] Set $\bar{\boldsymbol{R}}_{n} = \boldsymbol{R}_{n}^{(m)}$, $\forall n$, and update $\boldsymbol{Q}^{(m)}_{n}$, $\forall n$, as the solution to Problem (P2.2) obtained via Algorithm \ref{table2};
\end{itemize}
\item[3.] Until convergence.
\end{itemize}
\vspace{0.2cm} \hrule \label{table3}
\end{center}
\end{table}



\subsection{Computational Complexity Analysis}\label{Sec:Alg3_Complexity}

In this subsection, we provide the computational complexity analysis of the proposed algorithm for joint optimization of transmission covariance matrices and compression noise covariance matrices.

The computational complexity of proposed algorithm mainly comes from three parts. The first part is the derivation of the PFIM in (\ref{equ:F_0xi}) and the convex surrogate functions in (\ref{equ:relax_FC_optR}) and (\ref{equ:relax_FC_optT}), which consist of a lot of integral calculations. Let $\mathcal{O}(\Theta)$ denote the complexity of the integral of a function $f(\boldsymbol{q})$ over $\boldsymbol{q}$. Therefore, the complexity in this part can be given as $\mathcal{O}\left((4NK^2(M_{t}^2+M_{r}^2)+2NI_{A}I_{S})\Theta\right)$ where $I_{A}$ and $I_{S}$ denotes the average number of iterations for the AO and the SCA-based optimization. Then we analyze the computational complexities of the other two parts of transmit covariance matrices optimization and the compression noise variance matrices optimization. For the former, the relaxed problem (\ref{opt:relax_P2.1}) in each iteration have $2(N+K)$ constraints, $M_t^2$ complex variables, and $2K$ real variables, the computational complexity can be given as $\mathcal{O}\left( (2M_t^2+2K)^{3.5} + 2(N+K)(2M_t^2+2K)^{2.5} \right)$. For the latter, in each iteration, the computational complexity of solving problem (\ref{eqn:relax_P2.2}) is $\mathcal{O}\left( (2M_r^2+2K)^{3.5} + 2(N+K)(2M_r^2+2K)^{2.5} \right)$. At last, the whole computational complexity of Algorithm III can be given by $\mathcal{O}\big( 4NK^2(M_{t}^2+M_{r}^2)\Theta + I_{A}I_{S}(2N\Theta$ $+ 2(N+K)(2M_t^2+2K)^{2.5} + (2M_t^2+2K)^{3.5} + 2(N+K)(2M_r^2+2K)^{2.5} + (2M_r^2+2K)^{3.5}) \big)$.

\section{An Estimate-Then-Beamform-Then-Compress Strategy When $N_r$ is Large}\label{sec:EBC}

In this section, we study how to reduce the complexity to design compression covariance matrices when the BS is equipped with a large number of receive antennas. A key observation from our numerical results (e.g., in Section \ref{sec:numerical_results}) is that when $M_r$ is very large, many eigenvalues of the compression covariance matrices obtained via Algorithm III are extremely large. This indicates that when the dimension of the received signal is large, the BS will first transform the received signal into a low-dimension space and then perform quantization at these reduced dimensions. This is not surprising. When $M_r$ is large, if the quantization bits are allocated to quantize the signals received by all antennas, then the error of quantizing each signal is very large, leading to poor localization performance at the CU.

When $M_r$ is large, although at last only a low-dimension signal is quantized, at the beginning, we need to optimize the compression covariance matrices over a $M_r$-dimension space to find how to reduce the dimension of the received signal. As a result, as shown in Section IV-C, the complexity of Algorithm III is proportional to $M_r^7$. In this section, we will propose a low-complexity EBC strategy, under which the BS first estimates the AOAs of the signals from the targets (AOA estimation is quite accurate when $M_r$ is large), then beamforms the received signals into a low-dimension space, each pointing to one AOA direction, and last compresses the low-dimension signal and sends the quantization bits to the CU. Because compression design is over a lower dimension, the complexity of this scheme will be significantly reduced.

\subsection{Receive Beamforming and Compression Design}

Under our proposed estimate-then-beamform-then-compress strategy, the BS first estimates the AOAs, i.e., $\boldsymbol{\theta}_n, n = 1,\dots,N$, based on the received signal $(\ref{equ:signal_model})$ via the advanced AOA estimation method, e.g. MUSIC algorithm. Let $\hat{\boldsymbol{\theta}}_n, n = 1,\dots,N$ denote the AOA estimations. In the regime with large $M_r$, AOA estimation can be quite accurate. Then, the BS actively reduce the dimension of its received signals from $M_r$ to $L_r \le M_r$ via beamforming
\begin{align}\label{equ:y_b}
    \dot{\boldsymbol{y}}_{n} &= \boldsymbol{C}_n^H\boldsymbol{y} = \boldsymbol{C}_n^H \boldsymbol{A}_{n}\sum_{u=1}^{N}  \boldsymbol{B}_{n,u} \boldsymbol{V}_{u}^T \boldsymbol{x}_{u} + \boldsymbol{C}_n^H \boldsymbol{z}_{n}, ~\forall n,
\end{align}
where $\boldsymbol{C}_n \in \mathbb{C}^{M_r \times L_r}$ is the receive beamforming matrix at BS $n$. 
Then we also consider the Gaussian test channel for the compression process on $\dot{\boldsymbol{y}}_{n}$ as
\begin{align}\label{equ:y_bq}
    \ddot{\boldsymbol{y}}_{n} = \dot{\boldsymbol{y}}_{n} + \ddot{\boldsymbol{e}}_n, \forall n,
\end{align}
where $\ddot{\boldsymbol{e}}_n \sim \mathcal{CN}(\boldsymbol{0},\ddot{\boldsymbol{Q}}_{n}) \in \boldsymbol{C}^{L_r \times 1}$ is the Gaussian quantization error vector with zero mean and covariance $\ddot{\boldsymbol{Q}}_{n} \in \mathbb{S}^{L_r}_{++}$ at BS $n$. Based on the signal model in (\ref{equ:y_bq}), the fronthaul rate can be given by
\begin{align}\label{equ:cons_fronthaul_bq}
    & D_n(\{\boldsymbol{R}_n\}_{n=1}^N,\boldsymbol{C}_n,\ddot{\boldsymbol{Q}}_n) =  \nonumber \\ 
    & \mathbb{E}_{\boldsymbol{q}} \left[\log_2\left|\boldsymbol{C}_n^H\left(\sum\limits_{u=1}^N\boldsymbol{G}_{n,u}\boldsymbol{R}_n\boldsymbol{G}_{n,u}^H\right)\boldsymbol{C}_n + \sigma^2\boldsymbol{I}+\ddot{\boldsymbol{Q}}_{n}\right|\right] \notag \\
    & - \log_2\left|\ddot{\boldsymbol{Q}}_{n}\right|, ~\forall n.
\end{align}Moreover, the PFIM $\ddot{\boldsymbol{F}}_{\boldsymbol{\xi}}$ can be modeled as
\begin{align}\label{equ:PFIM_EBC}
    \ddot{\boldsymbol{F}}_{\boldsymbol{\xi}} = \ddot{\boldsymbol{F}}_{0,\boldsymbol{\xi}} + \ddot{\boldsymbol{F}}_{p},
\end{align}
where $\ddot{\boldsymbol{F}}_{p}$ is similarly to (\ref{eqn:dis}) with the refined target location distribution $\ddot{p}(\boldsymbol{q})$ based on the estimation, and the FIM is
\begin{align}\label{equ:F_0zeta_C}
    \ddot{\boldsymbol{F}}_{0,\boldsymbol{\zeta}} = \mathbb{E}_{\ddot{\boldsymbol{y}}} \left[ \frac{\partial \log p(\ddot{\boldsymbol{y}}|\boldsymbol{\zeta})}{\partial \boldsymbol{\zeta}} \Bigg(\frac{\partial \log p(\ddot{\boldsymbol{y}}|\boldsymbol{\zeta})}{\partial \boldsymbol{\zeta}}\Bigg)^T \right],
\end{align}
where $\ddot{\boldsymbol{y}} = [\ddot{\boldsymbol{y}}_1,\dots,\ddot{\boldsymbol{y}}_{N}]^T$ is the aggregated received signals under the proposed EBC strategy.
Based on the same approach used in Proposition \ref{prop:FIM_MC}, by simply changing $\boldsymbol{A}_n$ to $\boldsymbol{C}_n^H\boldsymbol{A}_n$, $\dot{\boldsymbol{A}}_n$ to $\boldsymbol{C}_n^H\dot{\boldsymbol{A}}_n$, and $\boldsymbol{O}_n$ to $\ddot{\boldsymbol{O}}_n = \sigma^2\boldsymbol{C}_n^H\boldsymbol{C}_n + \ddot{\boldsymbol{Q}}_n$,
the FIM $\ddot{\boldsymbol{F}}_{0,\boldsymbol{\zeta}}$ can be obtained by the following proposition.
\begin{proposition}\label{Prop:FIM_FIMC}
The FIM $\ddot{\boldsymbol{F}}_{0,\boldsymbol{\zeta}}$ characterized in (\ref{equ:F_0zeta_C}) can be expressed as
\begin{align}
    \ddot{\boldsymbol{F}}_{0,\boldsymbol{\zeta}} = 2\begin{bmatrix}
    \Re\{\ddot{\boldsymbol{F}}_1\} & \Re\{\ddot{\boldsymbol{F}}_2\} & -\Im\{\ddot{\boldsymbol{F}}_2\} \\
    \Re\{\ddot{\boldsymbol{F}}_2\}^T & \Re\{\ddot{\boldsymbol{F}}_3\} & -\Im\{\ddot{\boldsymbol{F}}_3\} \\
    -\Im\{\ddot{\boldsymbol{F}}_2\}^T & -\Im\{\ddot{\boldsymbol{F}}_3\}^T & \Re\{\ddot{\boldsymbol{F}}_3\}
    \end{bmatrix},
\end{align}
where $\ddot{\boldsymbol{F}}_1$, $\ddot{\boldsymbol{F}}_2$, and $\ddot{\boldsymbol{F}}_3$ have the similar definition to (\ref{equ:F1}) -- (\ref{equ:F3}), just with the terms $\left((\boldsymbol{A}_i^l)^H\boldsymbol{O}^{-1}_i\boldsymbol{A}_i^r\right)$ and replaced by $\Big((\boldsymbol{A}_i^l)^H\boldsymbol{C}_i^H\ddot{\boldsymbol{O}}^{-1}_i\boldsymbol{C}_i\boldsymbol{A}_i^r\Big)$, for $i = 1,\dots,N$, where $\boldsymbol{A}_i^l, \boldsymbol{A}_i^r \in \Big\{\boldsymbol{A}_i, \dot{\boldsymbol{A}}_i\Big\}$.
\end{proposition}

It is obvious that the receive beamforming matrices $\{\boldsymbol{C}_n\}_{n=1}^{N}$ also affect the FIM $\ddot{\boldsymbol{F}}_{0,\boldsymbol{\zeta}}$ via Proposition \ref{Prop:FIM_FIMC}. In the special case with $\boldsymbol{C}_{n} = \boldsymbol{I}, \forall n$, the FIM can be reduced to the one given by Proposition \ref{prop:FIM_MC}. Intuitively, utilizing $L_r < M_r$ beamformers at each BS can deteriorate the sensing performance even if the fronthaul capacity is unlimited. However, we can show that the estimation performance loss can be possibly avoided under appropriate receive beamforming design.

\begin{theorem}\label{thm:thm2}
In the case where the fronthaul capacity is unlimited such that there is no compression noise at each BS $n$, i.e., $\boldsymbol{Q}_n = \boldsymbol{0}, \forall n$, there is no estimation performance loss if the receive beamforming matrix satisfies
\begin{align}\label{equ:C_thm2}
    \mathcal{R}\{\boldsymbol{\Delta}_n(\boldsymbol{\theta}_n)\} \subseteq \mathcal{R}\{\boldsymbol{C}_n\}, \quad \forall n,
\end{align}
where $\boldsymbol{\Delta}_n(\boldsymbol{\theta}_n) = [\boldsymbol{A}_n(\boldsymbol{\theta}_n),\boldsymbol{\Pi}^{\perp}_{\boldsymbol{A}(\boldsymbol{\theta}_n)}\dot{\boldsymbol{A}}(\boldsymbol{\theta}_n)] \in \mathbb{C}^{M_t \times 2K}$ with $\boldsymbol{\Pi}^{\perp}_{\boldsymbol{A}(\boldsymbol{\theta}_n)} = \boldsymbol{I} - \boldsymbol{A}(\boldsymbol{\theta}_n)\left(\boldsymbol{A}(\boldsymbol{\theta}_n)^H\boldsymbol{A}(\boldsymbol{\theta}_n)\right)^{-1}\boldsymbol{A}(\boldsymbol{\theta}_n)^H \in \mathbb{C}^{M_t \times M_t}$.
\end{theorem}
\begin{IEEEproof}
    Please refer to Appendix \ref{app:thm2}
\end{IEEEproof}
By Theorem \ref{thm:thm2}, it illustrates that the receive beamforming is comprised by two parts of $\boldsymbol{A}_n(\boldsymbol{\theta}_n)$ and $\boldsymbol{\Pi}^{\perp}_{\boldsymbol{A}(\boldsymbol{\theta}_n)}\dot{\boldsymbol{A}}(\boldsymbol{\theta}_n)$. The roles of these two parts can be intuitively understood as: beamformers in the column space of $\boldsymbol{\Pi}^{\perp}_{\boldsymbol{A}(\boldsymbol{\theta}_n)}\dot{\boldsymbol{A}}(\boldsymbol{\theta}_n)$ are essential to the target location determination in terms of AOA $\boldsymbol{\theta}_n$; while beamformers in the column space of $\boldsymbol{A}(\boldsymbol{\theta}_n)$ are vital to preserve the information of the unknown attenuation coefficients $\boldsymbol{b}_{n}$, which also affects the localization performance.

Theorem \ref{thm:thm2} indicates that when $M_r$ is large, each BS can locally reduce the $M_r$-dimension signal to a $2K$-dimension signal without information loss in the ideal case with infinite fronthaul capacity. Note that the estimated AOAs $\hat{\boldsymbol{\theta}}_n$'s are very close to the true AOAs $\boldsymbol{\theta}_n$'s when $M_r$ is large. 
Therefore, we propose to design the beamforming vector of BS $n$ as
\begin{align}
    \boldsymbol{C}_n = \boldsymbol{E}_r(\hat{\boldsymbol{\theta}}_n) \in \mathbb{C}^{M_r \times 2K}, ~\forall n,
\end{align}
where columns of $\boldsymbol{E}_r(\hat{\boldsymbol{\theta}}_n)$ are the $L_r = 2K$ eigenvectors corresponding to the $2K$ non-zeros eigenvalues of the matrix $\boldsymbol{\Delta}_n(\hat{\boldsymbol{\theta}}_n)\boldsymbol{\Delta}_{n}^H(\hat{\boldsymbol{\theta}}_n)$.

Then we can also formulate the joint transmission and compression covariance matrices optimization problem as
\begin{subequations}
\begin{align}
    (\text{P3})\min_{\{\ddot{\boldsymbol{R}}_n,\ddot{\boldsymbol{Q}}_n\}_{n=1}^N} &~ \sum_{i=1}^{2K} [(\ddot{\boldsymbol{F}}_{\boldsymbol{\xi}}(\{\ddot{\boldsymbol{R}}_n,\ddot{\boldsymbol{Q}}_n\}_{n=1}^N))^{-1}](i,i) \\
    \text{s.t.}~~~~~ &~ \text{tr}(\ddot{\boldsymbol{R}}_n) \le \bar{P}_n, ~ \forall n, \label{eqn:1_EBC}\\
    \quad &~ D_n(\{\ddot{\boldsymbol{R}}_n\}_{n=1}^N,\ddot{\boldsymbol{Q}}_n)\leq \bar{D}_n, ~ \forall n, \label{eqn:2_EBC} \\ 
    \quad &~ \ddot{\boldsymbol{Q}}_{n} \succeq \boldsymbol{0}, ~ \forall n, \label{eqn:3_EBC}
\end{align}
\end{subequations}
It is obvious that Problem (P3) is non-convex and we cannot directly solve it. However, we can easily find that Problem (P3) is very similar to Problem (P1), where only the dimension of variables are reduced. Thus, we can follow Section IV to solve Problem (P3). Specifically, we first employ Schur complement technique to transform Problem (P3) as
\begin{subequations}
\begin{align}
    (\text{P4})\min_{\{\ddot{\boldsymbol{R}}_n,\ddot{\boldsymbol{Q}}_n\}_{n=1}^N,\{t_i\}_{i=1}^{2K}} &~ \sum_{i=1}^{2K} t_i \\
    \text{s.t.}~~~~~~~~~ &~ \begin{bmatrix}
        \ddot{\boldsymbol{F}}_{\boldsymbol{\xi}}(\{\ddot{\boldsymbol{R}}_n,\ddot{\boldsymbol{Q}}_n\}_{n=1}^N) & \boldsymbol{e}_i \\
        \boldsymbol{e}_i^T & t_i
    \end{bmatrix} \succeq \boldsymbol{0}, ~ \forall i, \label{equ:cons_F_CRB} \\
    \quad &~ {\rm (\ref{eqn:1_EBC})-(\ref{eqn:3_EBC})}. \nonumber
\end{align}
\end{subequations}
Then, similarly, we also adopt the AO-based method to solve Problem (P4), which follows the procedure of Algorithm III to iteratively optimize $\ddot{\boldsymbol{R}}_n$'s and $\ddot{\boldsymbol{O}}_n$'s. It is also guaranteed that the objective function of Problem (P4) will be decreased after each iteration. Here, the two subproblems in each iteration of the AO-based method are solved by following Section \ref{sec:opt_Rn} and \ref{sec:opt_On}, respectively, whose details are omitted.

\subsection{Computational Complexity Analysis}
Here, we also give the computational complexity for the joint transmission and compression optimization problem under the proposed hybrid estimation and compression strategy. Under the proposed EBC strategy, the subproblem for optimizing compression noise covariance matrices has $L_r^2$ complex variables and $2K$ real variables. Therefore, the complexity can be expressed as $\mathcal{O}\big(4NK^2(M_{t}^2+M_{r}^2)\Theta + I_{A}I_{S}(2N\Theta+(2M_t^2+2K)^{3.5}+2(N+K)(2M_t^2+2K)^{2.5} + (2L_r+2K)^{3.5}) + 2(N+K)(2L_r+2K)^{2.5}\big)$.
Since $L_r \ll M_r$ in the massive MIMO scenario, the computational cost can be greatly saved by adopting the proposed EBC strategy.


\section{Numerical Results}\label{sec:numerical_results}

In this section, we provide numerical examples to verify our results. The systems consists of $N=2$ BSs unless otherwise specified, whose locations are $\boldsymbol{\rho}_1 = \left[\frac{\sqrt{3}}{2},0\right]^T$ km and $\boldsymbol{\rho}_2 = \left[-\frac{\sqrt{3}}{2},0\right]^T$ km, respectively, and multiple targets whose location distributions follow
\begin{align}\label{equ:loc_tar_dis}
    p_{\boldsymbol{q}_k}(\boldsymbol{q}_{k}) = \frac{1}{2\pi r_k^2}\exp\left(-\frac{||\boldsymbol{q}_{k}-\bar{\boldsymbol{q}}_{k}||_2^2}{2r_k^2}\right), ~\forall k,
\end{align}
where $\bar{\boldsymbol{q}}_{k}$ is the center of the possible location region of target $k$.
The power spectrum of the noise at the BSs is $-169$ dBm/Hz, and the channel bandwidth is $1$ MHz. We assume that the two BSs possess the same transmit power constraint, denoted by $\bar{P}=\bar{P}_1=\bar{P}_2$, and the same fronthaul capacity constraint, denoted by $\bar{D}=\bar{D}_1=\bar{D}_2$.

\begin{figure}[t]
    \centering
    \includegraphics[width=.45\textwidth]{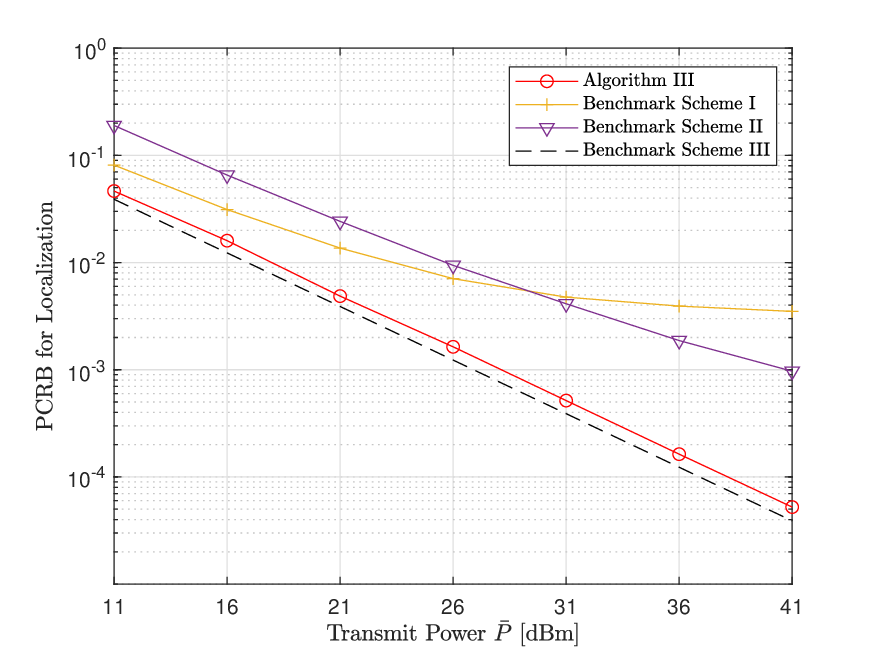}
    \caption{Sensing performance versus BSs' transmit power.}\vspace{-0.2cm}
    \label{fig:MSE_SNR}
\end{figure}

\begin{figure}[t]
    \centering
    \includegraphics[width=.45\textwidth]{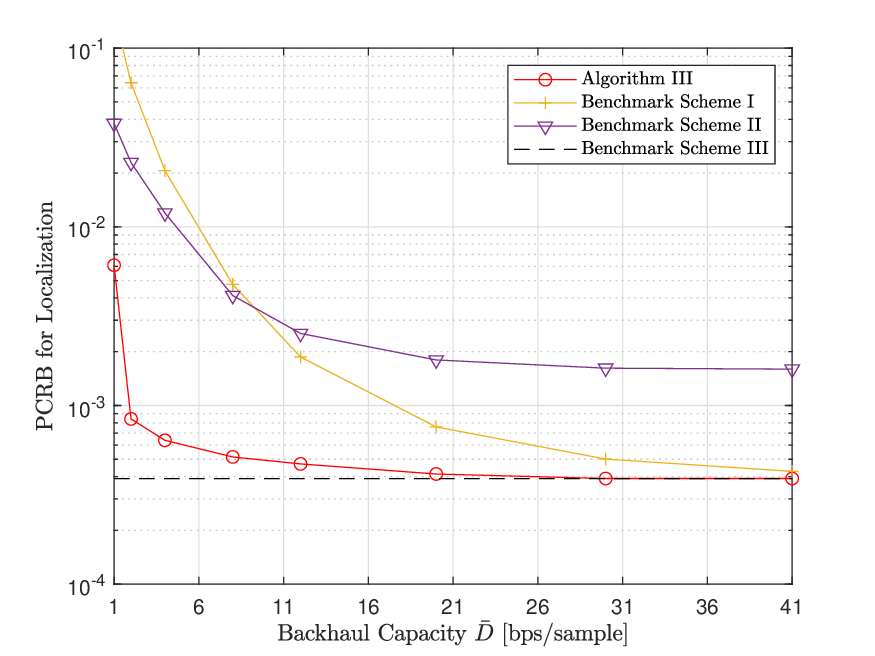}
    \caption{Sensing performance versus BSs' fronthaul capacity.}\vspace{-0.5cm}
    \label{fig:MSE_Capacity}
\end{figure}

\begin{figure}[t]
    \centering
    \includegraphics[width=.45\textwidth]{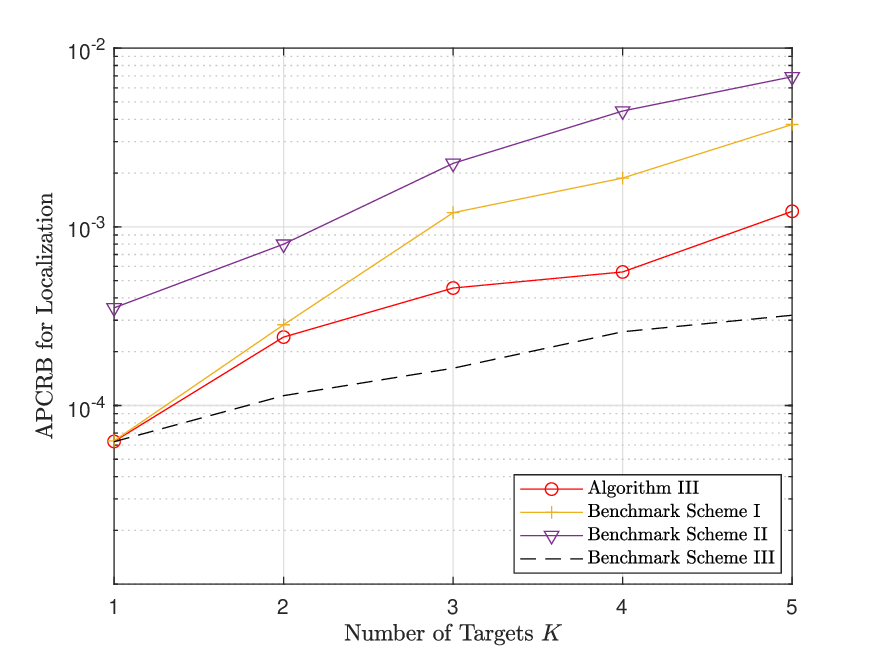}
    \caption{Sensing performance versus targets' number $K$.}\vspace{-0.5cm}
    \label{fig:MSE_K}
\end{figure}

\begin{figure}[t]
    \centering
    \includegraphics[width=.4\textwidth]{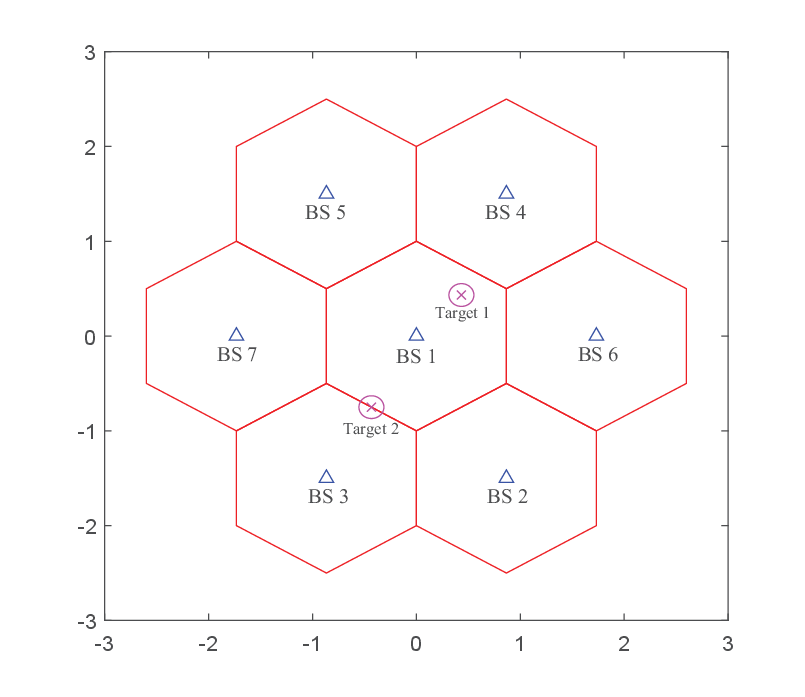}\vspace{-0.5cm}
    \caption{Diagram of the cellular system with $7$ hexagonal cells and target distributions.}
    \label{fig:Cellular}\vspace{-0.5cm}
\end{figure}

\begin{figure}[t]
    \centering
    \includegraphics[width=.45\textwidth]{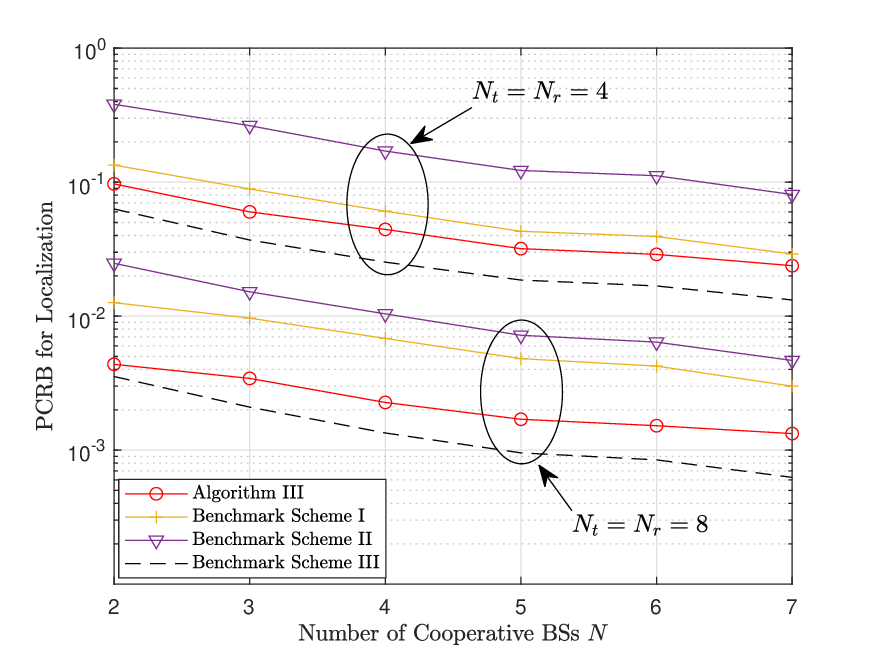}
    \caption{Sensing performance versus the number of cooperative BSs $N$.}
    \label{fig:MSE_NBS}\vspace{-0.5cm}
\end{figure}


\begin{figure}[t]
    \centering
    \includegraphics[width=.45\textwidth]{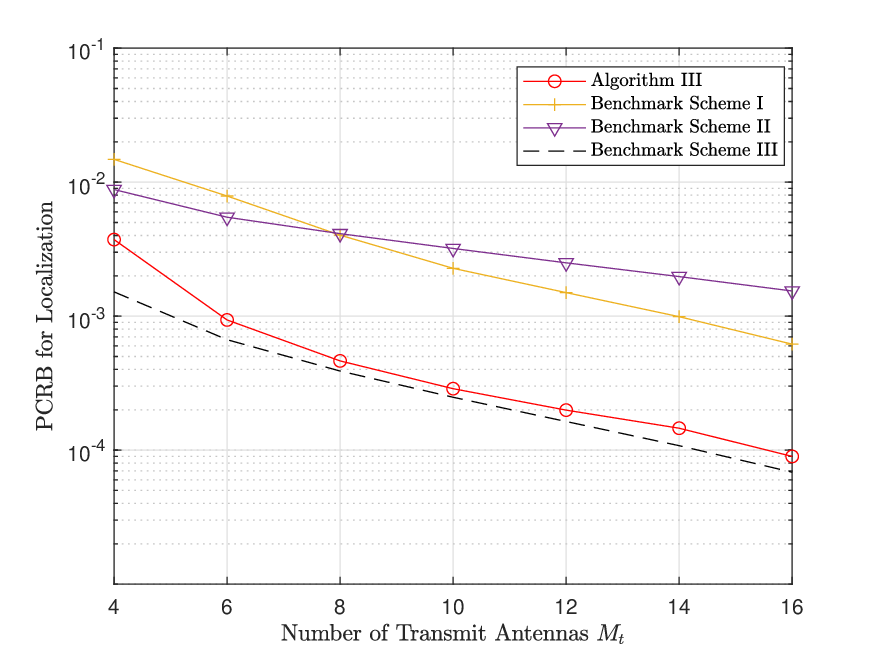}
    \caption{Sensing performance versus the number of transmit antennas $M_t$.}\vspace{-0.5cm}
    \label{fig:MSE_Nt}
\end{figure}

\begin{figure}[t]
    \centering
    \includegraphics[width=.45\textwidth]{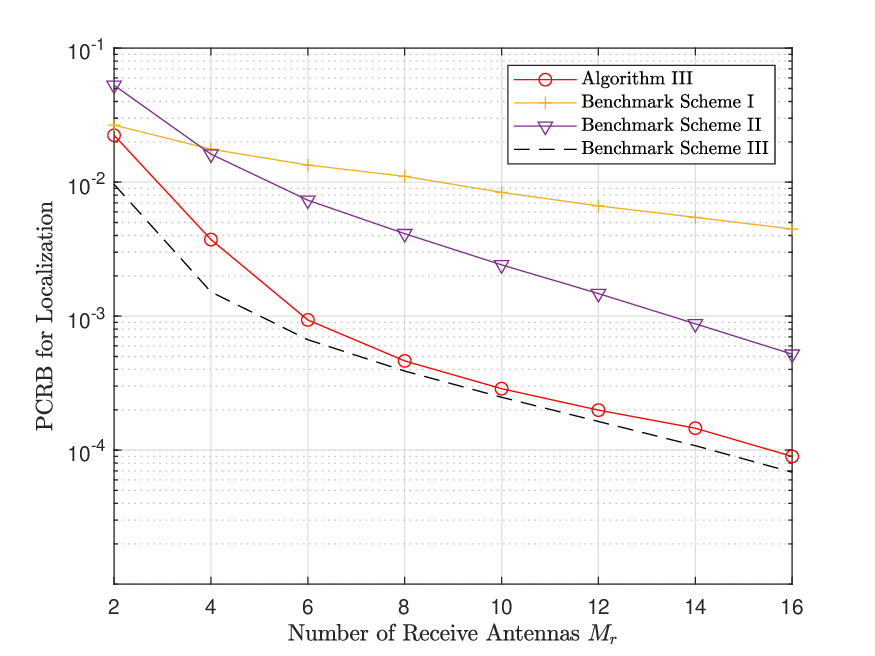}
    \caption{Sensing performance versus the number of receive antennas $M_r$.}
    \label{fig:MSE_Nr}
    \vspace{-0.5cm}
\end{figure}

\subsection{Performance Evaluation of Algorithm III}

First, we consider the performance of our proposed algorithm in a system with a general number of BS antennas. To verify the effectiveness of Algorithm \ref{table3} to solve Problem (P2), we provide several benchmark schemes: 
\begin{itemize}
    \item \textbf{Benchmark Scheme I}: we perform uniform quantization to the received signals of each BS receive antenna as shown in \cite{Liu_2015_TSP} and merely optimize the transmit covariance matrices based on Algorithm \ref{table1};
    \item \textbf{Benchmark Scheme II}: we fix the transmit covariance matrices of the BSs as $\boldsymbol{R}_1=\boldsymbol{R}_2=\frac{\bar{P}}{M_t}\boldsymbol{I}$ and merely optimize the compression noise covariance matrices based on Algorithm \ref{table2};
    \item \textbf{Benchmark Scheme III}: we assume an ideal case where the fronthaul capacity of each BS is infinite, i.e., constraint (\ref{eqn:2}) can be ignored in Problem (P2) and we merely design the transmit covariance matrices. This scheme provides performance lower bound.
\end{itemize}

We start with a numerical example with two targets. Their location distribution follows (\ref{equ:loc_tar_dis}), 
with $\bar{\boldsymbol{q}}_1 = \left[\frac{\sqrt{3}}{4},\frac{3}{4}\right]^T \text{km}, r_1 = 0.03$ km and $\bar{\boldsymbol{q}}_2 = \left[0,\frac{3}{4}\right]^T \text{km}, r_2 = 0.048$ km. Moreover, we assume that $M_t = M_r = 8$.
Fig. \ref{fig:MSE_SNR} shows the PCRB performance achieved by various schemes versus BSs' transmit power, when the common fronthaul capacity is set as $\bar{D}=8$ bits per second per sample. It is observed that our proposed algorithm achieves much lower PCRB compared to Benchmark Schemes I and II under different transmit power by benefiting from the joint optimization, while almost the same PCRB as the ideal case achieved by Benchmark Scheme III. The results also show that Benchmark Scheme II gradually outperforms Benchmark Scheme I when the transmit power is large, since the compression noise turns to act as the main limiting factor in this case.

Moreover, Fig. \ref{fig:MSE_Capacity} shows the PCRB performance achieved by various schemes versus BSs' fronthaul capacity, when the common transmit power is set as $\bar{P}=31$ dBm. Similarly, it is observed that our proposed algorithm achieves much lower PCRB compared to Benchmark Schemes I and II when the fronthaul capacity is moderate. When the fronthaul capacity is sufficiently large, our proposed algorithm and Benchmark Schemes I and III achieve similar PCRB because compression methods have little impact on performance. 
In the case with extremely limited fronthaul capacity, Benchmark Scheme II greatly outperforms Benchmark Scheme I, since the useful echo signals can be illy polluted by the compression noise without elaborated compression design even if the transmission covariance matrices are well optimized.

We also investigate the performance of the proposed algorithm when the number $K$ of targets ranges from $1$ to $5$ in Fig. \ref{fig:MSE_K}. Among these possible $5$ targets, two targets' location distribution is specified in the above, while the location distribution of the other three targets follows (\ref{equ:loc_tar_dis}) with $\bar{\boldsymbol{q}}_3 = \left[-\frac{\sqrt{3}}{4},\frac{3}{4}\right]^T$ km, $\bar{\boldsymbol{q}}_4 = \left[\frac{\sqrt{3}}{4},\frac{\sqrt{3}}{4}\right]^T$ km, $\bar{\boldsymbol{q}}_5 = \left[-\frac{1}{4},\frac{1}{2}\right]^T$ km and $r_3 = 0.03$ km, $r_4 = 0.03$ km, $r_5 = 0.048$ km. We consider that $M_t = M_r = 16$ and $\bar{P} = 21$ dBm in the numerical results. For fair comparison, we define the average PCRB (APCRB) as the performance metric defined by $\text{APCRB} = \text{PCRB}/K$, which can be seen as the average estimation error for one target. It is observed that  Algorithm III can significantly outperform Benchmark Scheme I and II when $K$ increases. Considering that multiple targets usually exist in the multi-cell systems, the joint transmission and compression design can greatly boost the localization accuracy in networked sensing systems. In the special case with $K=1$, Benchmark Scheme I and Algorithm III almost achieve the same performance as the performance lower bound, which is due to the fact that $\bar{D} = 8$ bits is enough to convey the localization information of one target and we only need to optimize the transmit covariance matrices.

Next, we also evaluate the proposed algorithm in the practical cellular system, which consists of $7$ hexagonal cells as shown in Fig. \ref{fig:Cellular}. In each cell, there is a BS locating at the center and the distance between two adjacent BSs is $\sqrt{3}$ km.
Here, we also consider that $\bar{D} = 8$ bps/sample and two targets in the network. The PCRB performance under two cases with different setup on the number of antennas are given. With more BSs for cooperative target localization, both the PCRB performance of Algorithm III and Benchamrk III can be significantly enhanced. However, the computational complexity also increases when $N$ is enlarged. In practical implementations, the number $N$ should be carefully selected by considering the trade-off between the PCRB performance and the computational costs. 

We show the impact of the number of transmit antennas and receive antennas on the PCRB performance in Fig. \ref{fig:MSE_Nt} and Fig. \ref{fig:MSE_Nr} with $\bar{D} = 8$ bps/sample and $\bar{P}=31$ dBm, respectively. Consistently, the proposed joint transmission and compression optimization can approach the Benchmark III with $M_t$ or $M_r$ being enlarged. The results also implies that optimizing transmit covariance matrices gives larger performance gains when either $M_t$ is large or $M_r$ is small, while the compression noise variance matrices optimization shifts to provide more performance enhancement when $M_t$ is small or $M_r$ is large. 
In particular, when $M_r$ increases, we can observe that the PCRB of Benchmark II decreases greatly, which adopts omnidirectional transmission, i.e., $\boldsymbol{R}_{n} = \boldsymbol{I}, \forall n$. This also verifies our illustration that utilizing massive receive antennas can provide accurate AOA estimation in Section \ref{sec:EBC} and motivates the design of the EBC strategy.

\begin{figure}
  \centering
  \includegraphics[width=.45\textwidth]{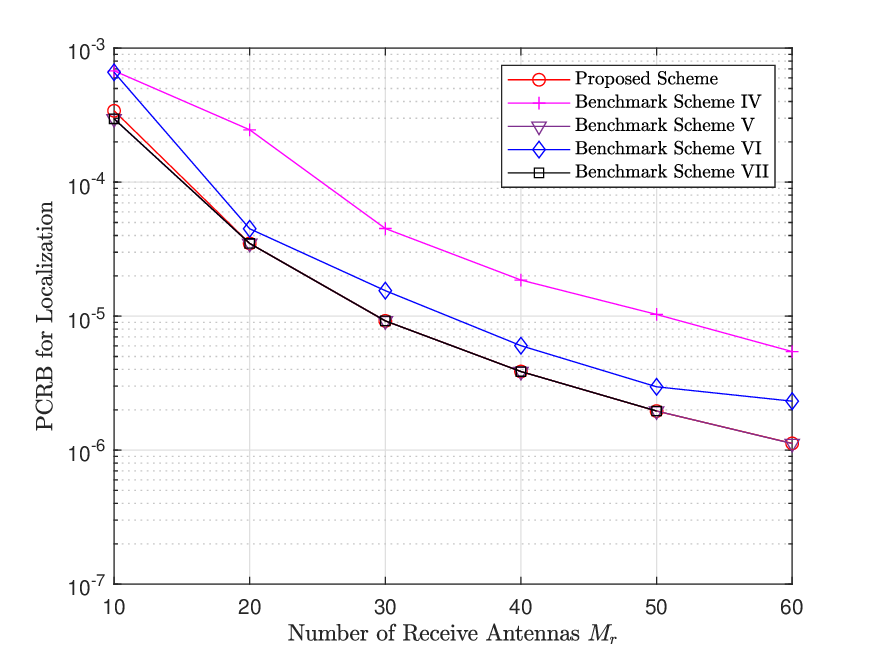}
  \caption{PCRB performance when $M_r$ is large.}\label{Fig:MSE_mMIMO}
  \vspace{-0.5cm}
\end{figure}

\begin{figure}
  \centering
  \includegraphics[width=.45\textwidth]{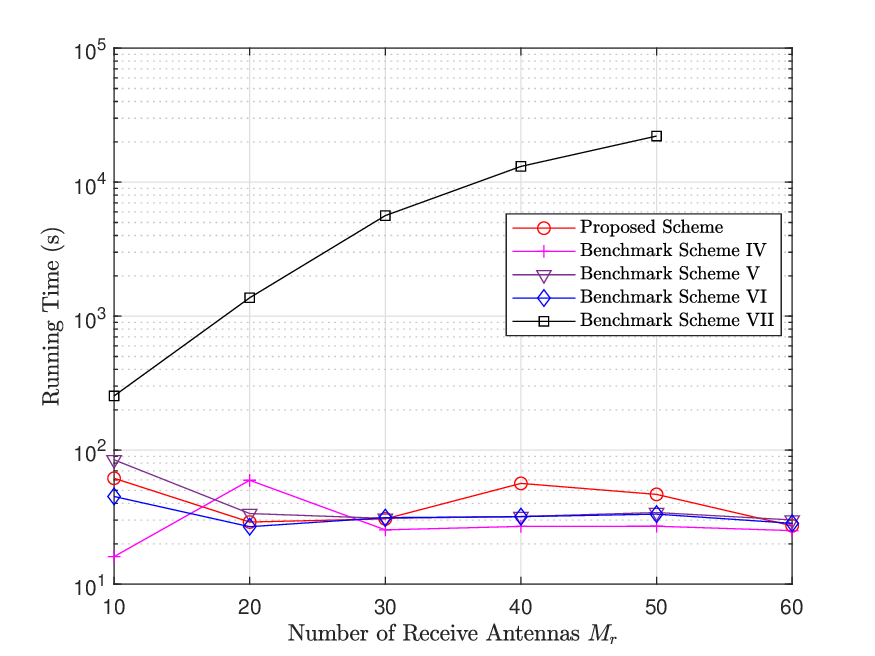}
  \caption{Running time cost when $M_r$ is large.}\label{Fig:RT_mMIMO}
  \vspace{-0.5cm}
\end{figure}

\subsection{Performance Evaluation of the Proposed EBC Strategy}

Finally, the performance of the proposed EBC strategy is evaluated when the BSs are equipped with a large number of receive antennas. 
Here, we assume $M_t = 8$ and $\bar{P} = 31$ dBm. Moreover, the distribution of two targets' locations is the same as that used in Fig. \ref{fig:MSE_SNR}. Besides the proposed scheme with $\boldsymbol{C}_n = \boldsymbol{E}_r(\hat{\boldsymbol{\theta}}_n)$ under EBC strategy, we also consider four benchmark schemes in this subsection:
\begin{itemize}
    \item \textbf{Benchmark Scheme IV}: under the EBC strategy, we set $L_r=2K-1$ and select columns from $\boldsymbol{E}_r(\hat{\boldsymbol{\theta}}_n)$ as the receive beamformers for each BS $n$.
    \item \textbf{Benchmark Scheme V}: under the EBC strategy, we set $L_r = 2K+1$, where the receive beamformers for each BS $n$ consist of all the $2K$ columns of $\boldsymbol{E}_r(\hat{\boldsymbol{\theta}}_n)$ and another one vector orthogonal to all columns of $\boldsymbol{E}_r(\hat{\boldsymbol{\theta}}_n)$.
    \item \textbf{Benchmark Scheme VI}: we select $L_r = 2K$ columns of the DFT matrix as receive beamformers and then optimize $\boldsymbol{R}_n$'s and $\ddot{\boldsymbol{O}}_n$'s jointly.
    \item \textbf{Benchmark Scheme VII}: we adopt $\boldsymbol{W}_n = \boldsymbol{I}, \boldsymbol{C}_n = \boldsymbol{I}, \forall n$ and thus the dimension of the received signal is not reduced before compression. This scheme can be considered as the performance lower bound for the EBC strategy.
\end{itemize}
Fig. \ref{Fig:MSE_mMIMO} and Fig. \ref{Fig:RT_mMIMO} show the performance and CPU running time of various schemes when the number of receive antennas ranges from $10$ and $60$. 
It is obvious that the proposed scheme has almost the same performance as Benchmark Scheme V and Benchmark Scheme VI, which verifies its effectiveness. 
We can only give the performance of Benchmark Scheme VI when $M_r \le 50$, since its computational complexity is prohibitively high that its performance is hard to be derived when $M_r > 50$. In contrast, the other four schemes can efficiently obtain the optimized transmission covariance matrices and compression noise covariance matrices even if $M_r$ is very large. The above results indicate that the proposed scheme under EBC strategy can achieve nearly optimal performance with computational efficiency and more suitable to be implemented in the massive MIMO scenario.

\section{Conclusions}

This work considered networked sensing in 6G cellular network, where multiple BSs forward their received echo signals to the CU via fronthaul links with limited capacity for jointly localizing multiple targets. We characterized the PCRB for the sensing performance in this setup, and designed an efficient algorithm to optimize the BS transmit covariance matrices and compression noise covariance matrices for minimizing this PCRB subject to fronthaul capacity constraints. Moreover, we proposed an estimate-then-beamform-then-compress strategy to realize near-optimal sensing performance with significant computational complexity reduction, when the BSs are equipped with a large number of receive antennas. Numerical results were provided to verify the effectiveness of our proposed schemes.

\begin{appendices}
\section{Proof of Proposition \ref{prop:FIM_MC}}\label{app:proA}

Based on the signal model in (\ref{eqn:compression}), each element in the FIM $\boldsymbol{F}_{0,\boldsymbol{\zeta}}$ in (\ref{eqn:zeta1}) can be calculated as
\begin{align}
    &\boldsymbol{F}_{0,\boldsymbol{\zeta}}(i,n) = \notag \\
    &2\Re\Bigg\{\sum_{m=1}^{N}\Bigg( \frac{\partial\big(\sum_{v=1}^{N}\boldsymbol{G}_{m,v}\boldsymbol{x}_v\big)^H}{\partial \zeta_{i}} \boldsymbol{O}_{m}^{-1} \frac{\partial\big(\sum_{v=1}^{N}\boldsymbol{G}_{m,v}\boldsymbol{x}_v\big)}{\partial \zeta_{n}} \Bigg) \Bigg\}, \notag \\
    &\quad\quad\quad\quad\quad\quad\quad\quad\quad\quad\quad\quad\quad\quad\quad\quad\quad\quad\quad \forall i, n.
\end{align}
where $\zeta_{i}$ and $\zeta_{j}$ are the $i$th element and $j$th element of $\boldsymbol{\zeta}$, respectively. Based on the observation that
\begin{align}
    \frac{\partial\boldsymbol{B}_{m,v}}{\partial b^{\rm R}_{u,w,k}} = -j\frac{\partial\boldsymbol{B}_{m,v}}{\partial b^{\rm I}_{u,w,k}}, ~\forall m,v,u,w,k,
\end{align}
we can obtain
\begin{align}
    \frac{\partial\big(\boldsymbol{G}_{m,v}\boldsymbol{x}_v\big)}{\partial b^{\rm R}_{u,w,k}} = -j\frac{\partial\big(\boldsymbol{G}_{m,v}\boldsymbol{x}_v\big)}{\partial b^{\rm I}_{u,w,k}}, ~\forall m,v,u,w,k.
\end{align}
As well, note that
\begin{align}
    \boldsymbol{F}_{0,\boldsymbol{\zeta}} &= \begin{bmatrix}
        \boldsymbol{F}_{\boldsymbol{\theta},\boldsymbol{\theta}} & \boldsymbol{F}_{\boldsymbol{\theta},\boldsymbol{b}^{\rm R}} & \boldsymbol{F}_{\boldsymbol{\theta},\boldsymbol{b}^{\rm I}} \\
        \boldsymbol{F}_{\boldsymbol{b}^{\rm R},\boldsymbol{\theta}} & \boldsymbol{F}_{\boldsymbol{b}^{\rm R},\boldsymbol{b}^{\rm R}} & \boldsymbol{F}_{\boldsymbol{b}^{\rm R},\boldsymbol{b}^{\rm I}} \\
        \boldsymbol{F}_{\boldsymbol{b}^{\rm I},\boldsymbol{\theta}} & \boldsymbol{F}_{\boldsymbol{b}^{\rm I},\boldsymbol{b}^{\rm R}} & \boldsymbol{F}_{\boldsymbol{b}^{\rm I},\boldsymbol{b}^{\rm I}}
    \end{bmatrix},
\end{align}
the FIM $\boldsymbol{F}_{0,\boldsymbol{\zeta}}$ can be re-expressed as the equation in (\ref{equ:F_theta_rcs}) with the definition as
\begin{align}
    &[\boldsymbol{F}_1(n,u)](i,k) = \notag \\
    &\sum_{m=1}^{N}\Bigg( \frac{\partial\big(\sum_{v=1}^{N}\boldsymbol{G}_{m,v}\boldsymbol{x}_v\big)^H}{\partial \theta_{n,i}} \boldsymbol{O}_{m}^{-1} \frac{\partial\big(\sum_{v=1}^{N}\boldsymbol{G}_{m,v}\boldsymbol{x}_v\big)}{\partial \theta_{u,k}} \Bigg), \\
    &[\boldsymbol{F}_2(n,u,w)](i,k) = \notag \\
    &\sum_{m=1}^{N}\Bigg( \frac{\partial\big(\sum_{v=1}^{N}\boldsymbol{G}_{m,v}\boldsymbol{x}_v\big)^H}{\partial \theta_{n,i}} \boldsymbol{O}_{m}^{-1} \frac{\partial\big(\sum_{v=1}^{N}\boldsymbol{G}_{m,v}\boldsymbol{x}_v\big)}{\partial b^{\rm R}_{u,w,k}} \Bigg), \\
    &[\boldsymbol{F}_3(n,v,u,w)](i,k) = \notag \\
    &\sum_{m=1}^{N}\Bigg( \frac{\partial\big(\sum_{v=1}^{N}\boldsymbol{G}_{m,v}\boldsymbol{x}_v\big)^H}{\partial b^{\rm R}_{u,w,k}} \boldsymbol{O}_{m}^{-1} \frac{\partial\big(\sum_{v=1}^{N}\boldsymbol{G}_{m,v}\boldsymbol{x}_v\big)}{\partial b^{\rm R}_{u,w,k}} \Bigg), \label{equ:F3_nvuw_ik}
\end{align}

Here, we first calculate the element $[\boldsymbol{F}_{1}(n,u)](i,k)$ in the FIM.
we can then obtain the left hand side of $\boldsymbol{O}_{m}^{-1}$ as equation (\ref{equ:u_n}).
\begin{figure*}
\begin{align}\label{equ:u_n}
    \frac{\partial \big(\sum_{v=1}^{N}\boldsymbol{G}_{n,v}\boldsymbol{x}_v \big)}{\partial \theta_{n,i}} &= \sum_{v=1}^{N} \frac{\partial(\boldsymbol{G}_{n,v}\boldsymbol{x}_v)}{\partial \theta_{n,i}} = \left\{\begin{array}{ll}
       \sum_{v=1}^{N} \Big(\dot{\boldsymbol{A}}_{n}\boldsymbol{e}_{i}\boldsymbol{e}_{i}^{T}\boldsymbol{B}_{n,v}\boldsymbol{V}^T_{v}\boldsymbol{x}_{v} \Big) + \boldsymbol{A}_{n}\boldsymbol{B}_{n,n}\boldsymbol{e}_{i}\boldsymbol{e}_{i}^{T}\dot{\boldsymbol{V}}^T_{n} \boldsymbol{x}_{n}, & m = n, \\
       \boldsymbol{A}_{m}\boldsymbol{B}_{m,n}\boldsymbol{e}_{i}\boldsymbol{e}_{i}^{T}\dot{\boldsymbol{V}}^T_{n} \boldsymbol{x}_{n} , & m \ne n.
    \end{array}\right.
\end{align}
\hrule
\end{figure*}
Based on (\ref{equ:u_n}), we consider to derive $F_{\theta_{n,i},\theta_{u,k}}$ individually in two cases of $n = u$ and $n \ne u$. When $n = u$, we can directly obtain the equation (\ref{equ:FIM_tt_n=u}),
\begin{figure*}
\begin{align}\label{equ:FIM_tt_n=u}
    & \frac{\partial\big(\sum_{v=1}^{N}\boldsymbol{G}_{m,v}\boldsymbol{x}_v\big)^H}{\partial \theta_{n,i}} \boldsymbol{O}_{m}^{-1} \frac{\partial\big(\sum_{v=1}^{N}\boldsymbol{G}_{m,v}\boldsymbol{x}_v\big)}{\partial \theta_{u,k}} = 
    \left\{ \begin{array}{ll}
        \vartheta_{n,i,k}, & m = n, \\
        \boldsymbol{x}_{n}^H\dot{\boldsymbol{V}}_{n}^{*}\boldsymbol{e}_{i}\boldsymbol{e}_{i}^{T}\boldsymbol{B}_{m,n}^{*}\boldsymbol{A}_m^H \boldsymbol{O}_{m}^{-1} \boldsymbol{A}_{m}\boldsymbol{B}_{m,n}\boldsymbol{e}_{k}\boldsymbol{e}_{k}^{T}\dot{\boldsymbol{V}}^T_{n} \boldsymbol{x}_{n}, & m \ne n,
    \end{array}\right.
\end{align}    
\hrule
\end{figure*}
where 
\begin{align}\label{equ:vartheta_nik}
    &\vartheta_{n,i,k} = \notag \\
    & \sum_{v=1}^{N} \sum_{w=1}^{N} \Big( \boldsymbol{x}_{v}^H\boldsymbol{V}_{v}^{*}\boldsymbol{B}_{n,v}^{*}\boldsymbol{e}_{i}\boldsymbol{e}_{i}^{T}\dot{\boldsymbol{A}}_n^H \boldsymbol{O}_{n}^{-1} \dot{\boldsymbol{A}}_{n}\boldsymbol{e}_{k}\boldsymbol{e}_{k}^{T}\boldsymbol{B}_{n,w}\boldsymbol{V}^T_{w}\boldsymbol{x}_{w} \Big) \notag \\
    & + \sum_{v=1}^{N} \Big( \boldsymbol{x}_{v}^H\boldsymbol{V}_{v}^{*}\boldsymbol{B}_{n,v}^{*}\boldsymbol{e}_{i}\boldsymbol{e}_{i}^{T}\dot{\boldsymbol{A}}_n^H \boldsymbol{O}_{n}^{-1} \boldsymbol{A}_{n}\boldsymbol{B}_{n,n}\boldsymbol{e}_{k}\boldsymbol{e}_{k}^{T}\dot{\boldsymbol{V}}^T_{n} \boldsymbol{x}_{n} \Big) \notag \\
    & + \sum_{w=1}^{N} \Big( \boldsymbol{x}_{n}^H\dot{\boldsymbol{V}}_{n}^{*}\boldsymbol{e}_{i}\boldsymbol{e}_{i}^{T}\boldsymbol{B}_{n,n}^{*}\boldsymbol{A}_n^H \boldsymbol{O}_{n}^{-1} \dot{\boldsymbol{A}}_{n}\boldsymbol{e}_{k}\boldsymbol{e}_{k}^{T}\boldsymbol{B}_{n,w}\boldsymbol{V}^T_{w}\boldsymbol{x}_{w} \Big) \notag \\
    & + \boldsymbol{x}_{n}^H\dot{\boldsymbol{V}}_{n}^{*}\boldsymbol{e}_{i}\boldsymbol{e}_{i}^{T}\boldsymbol{B}_{n,n}^{*}\boldsymbol{A}_n^H \boldsymbol{O}_{n}^{-1} \boldsymbol{A}_{n}\boldsymbol{B}_{n,n}\boldsymbol{e}_{k}\boldsymbol{e}_{k}^{T}\dot{\boldsymbol{V}}^T_{n} \boldsymbol{x}_{n}.
\end{align}
While $n \ne u$, we have the equation (\ref{equ:FIM_tt_n!=u}),
\begin{figure*}
\begin{align}\label{equ:FIM_tt_n!=u}
    & \frac{\partial\big(\sum_{v=1}^{N}\boldsymbol{G}_{m,v}\boldsymbol{x}_v\big)^H}{\partial \theta_{n,i}} \boldsymbol{O}_{m}^{-1} \frac{\partial\big(\sum_{v=1}^{N}\boldsymbol{G}_{m,v}\boldsymbol{x}_v\big)}{\partial \theta_{u,k}} =  \left\{ \begin{array}{ll}
        \omega_{n,u,i,k}, & m = n, m \ne u, \\
        \alpha_{n,u,i,k}, & m \ne n, m = u, \\
        \boldsymbol{x}_{n}^H\dot{\boldsymbol{V}}_{n}^{*}\boldsymbol{e}_{i}\boldsymbol{e}_{i}^{T}\boldsymbol{B}_{m,n}^{*}\boldsymbol{A}_m^H \boldsymbol{O}_{m}^{-1} \boldsymbol{A}_{m}\boldsymbol{B}_{m,u}\boldsymbol{e}_{k}\boldsymbol{e}_{k}^{T}\dot{\boldsymbol{V}}^T_{u} \boldsymbol{x}_{u}, & m \ne n, m \ne u,
    \end{array}\right.
\end{align}    
\hrule
\end{figure*}
where
\begin{align}
    &\omega_{n,u,i,k} = \notag \\
    &\sum_{v=1}^{N} \Big(\boldsymbol{x}_{v}^H\boldsymbol{V}^{*}_{v}\boldsymbol{B}_{n,v}^{*}\boldsymbol{e}_{i}\boldsymbol{e}_{i}^{T}\dot{\boldsymbol{A}}_{n}^{H} \boldsymbol{O}_{n}^{-1} \boldsymbol{A}_{n}\boldsymbol{B}_{n,u}\boldsymbol{e}_{k}\boldsymbol{e}_{k}^{T}\dot{\boldsymbol{V}}^T_{u} \boldsymbol{x}_{u} \Big) \notag \\
    &+ \boldsymbol{x}_{n}^H\dot{\boldsymbol{V}}^{*}_{n}\boldsymbol{e}_{i}\boldsymbol{e}_{i}^{T}\boldsymbol{B}_{n,n}^{*}\boldsymbol{A}_{n}^{H}  \boldsymbol{O}_{n}^{-1} \boldsymbol{A}_{n}\boldsymbol{B}_{n,u}\boldsymbol{e}_{k}\boldsymbol{e}_{k}^{T}\dot{\boldsymbol{V}}^T_{u}\boldsymbol{x}_{u}, \label{equ:omega_nuik} \\
    &\alpha_{n,u,i,k} = \notag \\
    &\sum_{v=1}^{N} \Big(\boldsymbol{x}_{n}^H\dot{\boldsymbol{V}}^{*}_{n}\boldsymbol{e}_{i}\boldsymbol{e}_{i}^{T}\boldsymbol{B}_{u,n}^{*}\boldsymbol{A}_{u}^{H} \boldsymbol{O}_{u}^{-1} \dot{\boldsymbol{A}}_{u}\boldsymbol{e}_{k}\boldsymbol{e}_{k}^{T}\boldsymbol{B}_{u,v}\boldsymbol{V}^T_{v}\boldsymbol{x}_{v} \Big) \notag \\
    &+ \boldsymbol{x}_{n}^H\dot{\boldsymbol{V}}^{*}_{n}\boldsymbol{e}_{i}\boldsymbol{e}_{i}^{T}\boldsymbol{B}_{u,n}^{*}\boldsymbol{A}_{u}^{H} \boldsymbol{O}_{u}^{-1} \boldsymbol{A}_{u}\boldsymbol{B}_{u,u}\boldsymbol{e}_{k}\boldsymbol{e}_{k}^{T}\dot{\boldsymbol{V}}^T_{u}\boldsymbol{x}_{u}. \label{equ:alpha_nuik}
\end{align}

Due to the fact $\text{tr}(\boldsymbol{U}\boldsymbol{V}) = \text{tr}(\boldsymbol{V}\boldsymbol{U})$ and $\boldsymbol{B}_{m,n}$ is a diagonal matrix, we can make some manipulations to express the matrix product term in (\ref{equ:FIM_tt_n=u}) as
\begin{align}\label{equ:ele_FIM_re}       
    & \boldsymbol{x}_{n}^H\dot{\boldsymbol{V}}_{n}^{*}\boldsymbol{e}_{i}\boldsymbol{e}_{i}^{T}\boldsymbol{B}_{m,n}^{*}\boldsymbol{A}_m^H \boldsymbol{O}_{m}^{-1} \boldsymbol{A}_{m}\boldsymbol{B}_{m,n}\boldsymbol{e}_{k}\boldsymbol{e}_{k}^{T}\dot{\boldsymbol{V}}^T_{n} \boldsymbol{x}_{n} \notag \\
    =& \Big(\boldsymbol{e}_{i}^{T}\boldsymbol{A}_m^H \boldsymbol{O}_{m}^{-1} \boldsymbol{A}_{m}\boldsymbol{e}_{k}\Big) \Big(\boldsymbol{e}_{k}^{T}\boldsymbol{B}_{m,n}\dot{\boldsymbol{V}}^T_{n} \boldsymbol{x}_{n}\boldsymbol{x}_{n}^H\dot{\boldsymbol{V}}_{n}^{*}\boldsymbol{B}_{m,n}^{*}\boldsymbol{e}_{i}\Big) \notag \\
    =& \Big[ \boldsymbol{A}_m^H \boldsymbol{O}_{m}^{-1} \boldsymbol{A}_{m}\Big](i,k) \Big[ \boldsymbol{B}_{m,n}^{*}\dot{\boldsymbol{V}}_{n}^{H}\boldsymbol{R}_{n}^{*}\dot{\boldsymbol{V}}_{n}\boldsymbol{B}_{m,n}\Big](i,k), \notag \\
    &\quad\quad\quad\quad\quad\quad\quad\quad\quad\quad\quad\quad\quad\quad\quad\quad\quad\quad m \ne n.
\end{align}
Then the matrix product terms in (\ref{equ:vartheta_nik})--(\ref{equ:alpha_nuik}) can also be expressed in the similar forms to (\ref{equ:ele_FIM_re}). 
At last, the matrix form of $\boldsymbol{F}_1(n,u)$ and $\boldsymbol{F}_1(n,n)$ can be represented as (\ref{equ:F1_ii}) and (\ref{equ:F1_in}), respectively.

Next, we aim to derive the expression of $\boldsymbol{F}_{2}$.
First, we can obtain
\begin{align}
    \frac{\partial\big(\sum_{v=1}^{N}\boldsymbol{G}_{m,v}\boldsymbol{x}_v\big)}{\partial b^{\rm R}_{u,w,k}} =& \left\{\begin{array}{ll}
        \boldsymbol{A}_{u}\boldsymbol{e}_{k}\boldsymbol{e}_k^{T}\boldsymbol{V}_{w}^{T}\boldsymbol{x}_{w}, & m = u, \\
        \boldsymbol{0}, & m \ne u.
    \end{array}\right. \label{equ:du_bR} \\
    \frac{\partial\big(\sum_{v=1}^{N}\boldsymbol{G}_{m,v}\boldsymbol{x}_v\big)}{\partial b^{\rm I}_{u,w,k}} =& \left\{\begin{array}{ll}
        j\boldsymbol{A}_{u}\boldsymbol{e}_{k}\boldsymbol{e}_k^{T}\boldsymbol{V}_{w}^{T}\boldsymbol{x}_{w}, & m = u, \\
        \boldsymbol{0}, & m \ne u.
    \end{array}\right. \label{equ:du_bI}
\end{align}
Similar to the results in the above, the elements in $\boldsymbol{F}_2(n,u,w)$ can be given in (\ref{equ:F2_iii}) -- (\ref{equ:F2_ink})

Finally, based on (\ref{equ:F3_nvuw_ik}), (\ref{equ:du_bR}), and (\ref{equ:du_bI}), the elements in $[\boldsymbol{F}_3(n,v,u,w)](i,k)$ can be explicitly derived in (\ref{equ:F3_nini}) -- (\ref{equ:F3_niuk}).


\section{Proof of Theorem \ref{thm:thm2}}\label{app:thm2}

Based on the signal model (\ref{equ:y_b}), the covariance of the background noise $\dot{\boldsymbol{z}}_n = \boldsymbol{C}^H\boldsymbol{z}_n$ after performing receive beamforming can be given by $\dot{\boldsymbol{O}}_{n} = \sigma^2\boldsymbol{C}_n^H\boldsymbol{C}_n$.
Next, we should prove that the FIM with $\boldsymbol{C}_n = \boldsymbol{\Delta}_n(\boldsymbol{\theta}_n)\boldsymbol{L}_n$ is equal to the FIM with $\boldsymbol{C}_n = \boldsymbol{I}$, where $\boldsymbol{L}_n \in \mathbb{C}^{L \times L}$ is a non-singular matrix. According to Proposition \ref{Prop:FIM_FIMC}, we need to prove that the term 
\begin{align}
   (\boldsymbol{A}_{n}^l)^H\boldsymbol{C}_n\dot{\boldsymbol{O}}^{-1}_n\boldsymbol{C}_n^H\boldsymbol{A}_{n}^r = (\boldsymbol{A}_{n}^l)^H\boldsymbol{A}_{n}^r, ~\forall n, \label{equ:AQA_eq}
\end{align}
holds with any $\boldsymbol{A}_{n}^l, \boldsymbol{A}_{n}^r \in \left\{\boldsymbol{A}_n,\dot{\boldsymbol{A}}_n\right\}$.

We first consider the case where $\boldsymbol{A}_{n}^l = \boldsymbol{A}_n$ and $\boldsymbol{A}_{n}^r = \boldsymbol{A}_n$ as an example.
For the left-hand side of (\ref{equ:AQA_eq}), we have
\begin{align}\label{equ:prove_AQA_eq}
    &\boldsymbol{A}_{n}^H\boldsymbol{C}_n\dot{\boldsymbol{O}}^{-1}_n\boldsymbol{C}_n^H\boldsymbol{A}_n \notag \\
    =& \boldsymbol{A}_{n}^H [\boldsymbol{A}_{n},~ \boldsymbol{\Pi}_{\dot{\boldsymbol{A}}_n}^{\perp}\dot{\boldsymbol{A}}_{n}]\boldsymbol{L}_n \notag \\
    &\cdot\left\{\boldsymbol{L}_n^H\begin{bmatrix}
        \boldsymbol{A}_{n}^H\boldsymbol{A}_{n} & \boldsymbol{A}_{n}^H\boldsymbol{\Pi}_{\dot{\boldsymbol{A}}_n}^{\perp}\dot{\boldsymbol{A}}_{n} \\
        (\boldsymbol{\Pi}_{\dot{\boldsymbol{A}}_n}^{\perp}\dot{\boldsymbol{A}}_{n})^H\boldsymbol{A}_{n} & (\boldsymbol{\Pi}_{\dot{\boldsymbol{A}}_n}^{\perp}\dot{\boldsymbol{A}}_{n})^H\boldsymbol{\Pi}_{\dot{\boldsymbol{A}}_n}^{\perp}\dot{\boldsymbol{A}}_{n}
    \end{bmatrix}\boldsymbol{L}_n\right\}^{-1}  \notag \\
    &\cdot \boldsymbol{L}_n^H\begin{bmatrix}
        \boldsymbol{A}_n^H \\
        (\boldsymbol{\Pi}_{\dot{\boldsymbol{A}}_n}^{\perp}\boldsymbol{A}_{n})^H
    \end{bmatrix} \dot{\boldsymbol{A}}_n \notag \\
    =&\begin{bmatrix}
        \boldsymbol{A}_n^H\dot{\boldsymbol{A}}_n & \mathbf{0}
    \end{bmatrix} \begin{bmatrix}
        (\boldsymbol{A}_n^H\dot{\boldsymbol{A}}_n)^{-1} & \mathbf{0} \\
        \mathbf{0} & (\dot{\boldsymbol{A}}_{n}\Pi_{\dot{\boldsymbol{A}}_n}^{\perp}\dot{\boldsymbol{A}}_{n})^{-1}
    \end{bmatrix}\begin{bmatrix}
        \boldsymbol{A}_n^H\dot{\boldsymbol{A}}_n \\
        \mathbf{0}
    \end{bmatrix} \notag \\
    =& \boldsymbol{A}_n^H\boldsymbol{A}_n, ~\forall n.
\end{align}
Similarly, the equations in the other three cases can also be derived, which completes the proof. 

\end{appendices}

\bibliographystyle{IEEEtran}
\bibliography{reference}

\end{document}